\documentclass{aa}  

\usepackage{graphicx}
\usepackage{txfonts}
\usepackage{comment}
\usepackage{caption}
\usepackage{subfig}
\usepackage{lipsum}

\usepackage{booktabs}
\usepackage{array}

\newcommand{\abs}[1]{\lvert#1\rvert}

\newcommand{\taumax}{\tau_{\rm max}}
\newcommand{\worg}{w_{\rm org}}

\newcommand{\tausph}{\tau^{\rm sph}}

\newcommand{\sas}{S\&S }
\newcommand{\bsas}{S\&S}
\newcommand{\sps}{SPS }

\newcommand{\seps}{SePS }

\begin{document}

   \title{The scattering order problem in Monte Carlo radiative transfer}
        

   \author{A.~Krieger
          \and
          S.~Wolf
          }

   \institute{University of Kiel, Institute of Theoretical Physics and Astrophysics, Leibnizstrasse 15, 24118 Kiel, Germany, \\ \email{akrieger@astrophysik.uni-kiel.de}
}


   \date{Received: August 10, 2020; Accepted: November 5, 2020}


  \abstract
   {

   Radiative transfer simulation is an important tool that allows us to generate synthetic images of various astrophysical objects. In the case of complex three-dimensional geometries, a Monte Carlo-based method that  simulates photon packages as they move through and interact with their environment is often used. Previous studies have shown,  in the regime of high optical depths, that the required number of simulated photon packages strongly rises and estimated fluxes may be severely underestimated. In this paper we identify two problems that arise for Monte Carlo radiative transfer simulations that hinder a proper determination of flux:  first, a mismatch between the probability and weight of the path of a photon package and second, the necessity of simulating a wide range of high scattering orders. Furthermore, we argue that the peel-off method partly solves these problems, and we additionally propose an extended peel-off method. Our proposed method improves several shortcomings of its basic variant and relies on the utilization of precalculated sphere spectra. We then combine both peel-off methods with the Split method and the Stretch method and numerically evaluate their capabilities as opposed to the pure Split\,\&\,Stretch method in an infinite plane-parallel slab setup. We find that the peel-off method greatly enhances the performance of these simulations; in particular, at a transverse optical depth of $\taumax=75$ our method achieved a significantly lower error than previous methods while simultaneously saving ${>}95\%$ computation time. Finally, we discuss the inclusion of polarization and Mie-scattering in the extended peel-off method, and argue that it may be necessary to equip future Monte Carlo radiative transfer simulations with additional advanced pathfinding techniques.

   }


   \maketitle
   
%

\section{Introduction}  
   Radiative transfer (RT) simulations based on the Monte Carlo (MC) method play a leading role in the calculation of the radiative field and corresponding observable properties of a wide variety of astrophysical objects. They enable simulations of complex systems, such as  protoplanetary or circumbinary disks, Bok globules, filaments, or planetary atmospheres, which are subjects of studies of various Monte Carlo radiative transfer (MCRT) codes, for instance, RADMC-3D \citep{2012ascl.soft02015D}, POLARIS \citep{2016A&A...593A..87R}, and SKIRT\,9 \citep{2020A&C....3100381C}. The core principle of this MC approach is the simulation of photon packages that follow individual probabilistically determined paths through the model space in which they interact with matter, like dust or gas. This approach can then be used to calculate temperature distributions and flux maps of the simulated object.
   
   However, the computational demand for these simulations rises significantly with the optical depth of the system, and temperature and flux estimates may, despite this, show high levels of noise  \citep[e.g.,][]{2016A&A...590A..55B,2017A&A...603A.114G,2018ApJ...861...80C,2020A&A...635A.148K}. Several methods that aimed to solve these problems at high optical depths have been proposed, for instance, the partial diffusion approximation \citep{2009A&A...497..155M}, the modified random walk \citep{2009A&A...497..155M,2010A&A...520A..70R}, biasing techniques \citep{2016A&A...590A..55B}, and precalculated sphere spectra \citep{2020A&A...635A.148K}. However, despite the success of these methods in their respective areas of application, the overall problem of high optical depth has still not been solved.

   In general, these methods assist at least one of two types of MCRT simulations: the calculation of temperature distributions and/or flux maps. The aim of this paper is to study the latter in greater detail. \cite{2017A&A...603A.114G} conducted a study, in which the outcome of different RT codes for a simple three-dimensional setup were compared. The setup is  composed of a homogeneous slab of dust that is illuminated by a single blackbody source. Already at a transverse optical depth of $\taumax=30$ the accounted transmitted flux differences between these RT codes reached  ${\sim}100\%$. This was further investigated by \cite{2018ApJ...861...80C}, who called this the failure of MCRT at medium to high optical depths. In order to grasp the core problem that leads to this failure, they further simplified the setup to an infinite plane-parallel slab that is illuminated on one side and study the intensity of radiation that is transmitted through the slab. A comparison of the basic MC approach and two biasing techniques, the Split method and the Stretch method, showed the best results when both biasing techniques are used simultaneously. However, despite the relative success of the Split\,\&\,Stretch (\bsas) method, the equivalent serial computation time that was required to properly simulate a transverse barrier of $\taumax=75$ reached $52$\,days. The authors concluded that a proper simulation at high optical depths requires an enormous amount of simulated photon packages, and they suggested   equipping MCRT simulations with statistical methods to measure the convergence of their results. 
   
   In this study we  further investigate the core problem of MCRT simulations at high optical depths and demonstrate how the peel-off method greatly enhances their performance. In Sect. \ref{sec:setup_and_problem_analysis} we briefly describe the setup for our simulations, which is adopted from \cite{2018ApJ...861...80C}. We then examine and identify the main difficulty MCRT simulations face in the domain of high optical depths. Based on these results, we give reasons for the great success of the \sas method but also show its shortcomings. In Sect. \ref{sec:methods} we argue that the peel-off method bridges one of the core problems:  the need to calculate a broad range of scattering orders for a proper flux estimation. Additionally, we introduce the concept of the extended peel-off method, which may greatly boost the computation speed of a MCRT simulation, that is based on the basic peel-off method. Moreover, we describe how emission properties of spheres, which are crucial for this method, as well as their projections onto a detector can be precalculated in order to speed up MCRT simulations. We then argue why  using precalculated   emission spectra is preferred to using spectra based on the modified random walk \citep[MRW;][]{2009A&A...497..155M,2010A&A...520A..70R}. In Sect. \ref{sec:results_and_discussion} we apply the basic peel-off method as well as the extended peel-off method to the infinite slab setup, and compare their results to those of the \sas method. In this context we discuss the advantages and potential disadvantages of using  the peel-off method. In particular, we look into its capabilities of performing proper high optical depth simulations with the use of Mie-scattering and a full treatment of the polarization state of photon packages. Finally, in Sect. \ref{sec:summary} we briefly summarize our results. 

\section{Setup and problem identification}
\label{sec:setup_and_problem_analysis}

In this section the infinite plane-parallel homogeneous slab setup is described, and a MC based probabilistic numerical implementation and a non-probabilistic implementation are summarized. In Sect. \ref{sec:problems_high_optical_depth} two core problems of MCRT simulations at high optical depths are demonstrated on the basis of the slab experiment. Subsequently, the success of the \sas method at optical depths $\taumax<75$ and its failure at higher optical depths is explained, highlighting the need for more advanced MCRT methods, which will be discussed in detail later in Sect. \ref{sec:methods}.

\subsection{Setup: Infinite plane-parallel slab}
\label{sec:slab_numerical_implementation}
To study the performance and capabilities of MCRT simulations at high optical depths, we simulate an infinite plane-parallel homogeneous slab, which is illuminated on the top side by an isotropic monochromatic source of radiation. Our goal is to study this setup using various biasing techniques, which are described in  this section, and compare them to the methods described by \cite{2018ApJ...861...80C}. To ensure comparability, we adopt the  setup, i.e., a homogeneous slab with an albedo of $A=0.5$ under the assumption of isotropic scattering, and simulate the transmission through the slab. For the implementation we make use of optical depth units, which are the natural units of this problem, and simulate slabs of different transverse optical depths $\taumax$. Since this setup was described by \cite{2018ApJ...861...80C}, we narrow our description down to the key features.

   \begin{figure}
   \centering
   \includegraphics[width = 0.5\textwidth]{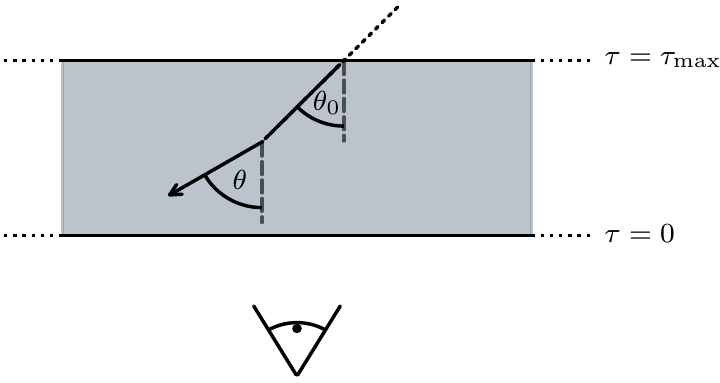}
   \caption{Sketch of the infinite plane-parallel slab setup. Radiation enters the slab of transverse optical depth $\tau=\taumax$ through the top layer. The observer detects radiation that is transmitted through the slab and is located below the bottom layer of the slab. The initial direction of a photon package is described by the penetration angle $\theta_0$. If the photon package scatters, its direction is chosen randomly according to its corresponding PDF and the new direction is described by the angle $\theta$.}
              \label{fig:sketch_slab_setup}%
   \end{figure} 

During the simulation $N$ photon packages are created that penetrate the top layer of the slab and travel through it, until they leave one side of the slab. The initial direction of any photon package is chosen randomly according to a probability density function (PDF) given by
\begin{equation}
p_{\mu_0}(\mu_0) = 2 \mu_0, \hfill  \text{with } 0\leq \mu_0 \leq 1,\hspace{1cm} 
\label{eq:pdf_mu_0}
\end{equation}
where $\mu_0 = \cos(\theta_0)$ is the cosine of the initial penetration angle $\theta_0$; in particular, $\mu_0=0$ and $\mu_0=1$ correspond to a direction parallel to the slab's surface layer and perpendicular to it, respectively. A sketch of the setup can be seen in Fig. \ref{fig:sketch_slab_setup}. Upon entry, a photon package travels a distance $\tau$ that is chosen randomly according to the PDF
\begin{equation}
p_\tau(\tau) = \exp\left( - \tau\right), \hfill \text{with } \tau\geq 0.\hspace{1cm}
\label{eq:pdf_tau}
\end{equation} 
At the resulting new position the photon package has a probability of absorption ($1-A$) and a probability of scattering ($A$). In the case of a scattering event, a new direction $\mu$ is assigned to the photon package.The PDF for isotropic scattering is given by
\begin{equation}
p_\mu(\mu) = \frac{1}{2}, \hfill \text{with } -1\leq \mu \leq 1. \hspace{1cm}
\label{eq:pdf_mu}
\end{equation}
The steps of translation, defined by Eq. \eqref{eq:pdf_tau}, and redirection, defined by Eq. \eqref{eq:pdf_mu}, of scattered photon packages are repeated until the translation distance $\tau$ of the photon package is large enough to leave the slab through one of its boundary layers. 
Photon packages that penetrate the bottom layer of the slab contribute to the transmission intensity. In order to estimate the contribution of a photon package to the transmitted intensity, it carries a weight $w$, which is modified during the MC procedure and eventually adds up at the detector to the overall transmitted intensity. During its creation, a photon package is neutral, meaning that its weight is initialized with $w=1$.
For a significant boost in performance we use the \sas method \citep{2018ApJ...861...80C}. When the Split method is used, the weight of the photon package is reduced for every event of interaction by the factor $w^{\text{split}}=A$. A photon package with reduced weight then continues to propagate through the slab. This biasing technique prohibits an early removal of the photon package due to absorption. When the Stretch method is used, the traversed path length is chosen according to the PDF for composite-biasing \citep{2016A&A...590A..55B}, which is given by
\begin{equation}
q_\tau(\tau) = \frac{1}{2}\left[ \exp\left( -\tau\right) + \alpha  \exp\left( -\alpha \tau \right)\right], \hfill \text{with } \alpha=\frac{1}{1+T}, \hspace{1cm}
\label{eq:pdf_q}
\end{equation}
where $T$ is the distance to the surface layer of the slab in the current direction of the photon package. In order to compensate for this modification of the path length, the weight of the photon packages needs to be readjusted by the factor 
\begin{equation}
w^{\text{stretch}} = \frac{p_\tau(\tau) }{q_\tau(\tau) }.
\end{equation}
The detector is defined such that transmitted radiation is sampled with respect to its direction of propagation (given by the variable $\mu$) when leaving the slab, resulting in a sampled directional transmission spectrum. The detector is composed of $M$ equal-sized bins for the directions $\mu$ that cover the interval $\left[ 0,1\right]$. The $j$-th detector bin of the detector measures the intensity $I_j$ given by
\begin{equation}
I_j = \frac{\sum_{n=1}^N w^{(n)}_j}{N} \frac{M}{2 \mu_j},
\end{equation}
where $ w^{(n)}_j$ is the contribution (i.e., weight) of the $n$-th photon package and $\mu_j$ the central value of the $j$-th detector bin.

\subsection{Non-probabilistic numerical implementation}
\label{sec:non-probabilistic_method}
The conceptually simple problem of determining the transmission of radiation through an infinite plane-parallel slab (see Sect. \ref{sec:slab_numerical_implementation}) has been studied extensively in the past. This setup is particularly interesting as it is challenging for the MC method, but at the same time solvable with a non-probabilistic method, making it a suitable testbed for MCRT simulations. Of the different approaches to solve this problem, the spherical harmonics method has shown to be particularly efficient \citep{2001MNRAS.326..722B}. For comparability, we adopt the method for solving this problem  described by \cite{2018ApJ...861...80C}, which is based on work done by  \cite{1983ApJ...275..292R} and its optimization proposed by \cite{1995MNRAS.277.1279D}. This method is based on a solution of the integral 
\begin{equation}
\mu \frac{\partial I}{\partial \tau}(\tau,\mu) = I(\tau,\mu) - \frac{A}{2} \int_{-1}^1 d\mu'\, I(\tau,\mu')\, \Psi(\mu,\mu'),
\end{equation}
where $\tau$ is the transverse optical depth and $\Psi(\mu,\mu')$ the normalized redistribution function for incoming ($\mu'$) and outgoing ($\mu$) radiation. Unless specifically stated otherwise, we assume an albedo of $A=0.5$ and isotropic scattering (i.e., $\Psi(\mu,\mu')=1$) throughout this paper. When assuming no additional internal sources, this differential equation is determined by its boundary conditions. In particular, the method relies on the series expansion of $I(\tau,\mu)$ in Legendre polynomials, which is truncated at some odd order $L$. The solution of $I(\tau,\mu)$ is then valid for $(L+1)/2$ directions $\mu$, which correspond to the zeros of the Legendre polynomial of order $L+1$. For comparability, we chose $L=81$. A more detailed description of this method can be found in the works of \cite{1983ApJ...275..292R}, \cite{1995MNRAS.277.1279D}, \cite{2001MNRAS.326..722B}, and \cite{2018ApJ...861...80C}. In later sections of this paper this method is referred to as the non-probabilistic solution to the infinite slab experiment.

\subsection{MCRT simulations at high optical depths}
\label{sec:problems_high_optical_depth}

High optical depths have been proven to cause major difficulties for MCRT simulations  \citep[e.g.,][]{2016A&A...590A..55B,2017A&A...603A.114G,2018ApJ...861...80C,2020A&A...635A.148K}. \cite{2018ApJ...861...80C} attempted to investigate the cause of this problem and find that at high transverse optical depths results of MCRT simulations strongly depend on the number of simulated photon packages. Additionally, the need for a theoretical understanding of this issue was highlighted and the use of statistical tests for convergence was recommended. The aims of this section aims are the deepening of our understanding of the core problem of MCRT simulations at high optical depths,  and thus  laying the foundation for a potential solution to it. To this end, we first give a brief description of flux determination in MCRT simulations before particularly analyzing the slab experiment.

Photon packages are emitted from one or more sources and subsequently follow their individual paths through the model space before eventually being detected. In order to determine flux accurately, in an ideal simulation, every possible path needs to be taken into account. Since this is not feasible in a realistic simulation, the number of simulated photon packages is limited. The complexity of potential paths increases with the number of interactions it undergoes. In other words, the volume of paths increases with the number of potential interactions. However, the contribution of a path to the measured intensity is highly dependent on its particular properties. In the MCRT simulation described in Sect. \ref{sec:slab_numerical_implementation} the contribution of a path is represented by the weight of the corresponding photon package. In order to determine a correct solution it is particularly important to simulate paths with high weights. To identify such paths it is mandatory to identify factors that reduce the weight of a path, i.e., that pose some form of resistance on the path of the photon package. In the case of the infinite plane-parallel slab, for instance, it is obvious that the paths of least resistance that contribute to detector bins of low $\mu$ values are those that  quickly reach the bottom layer of the slab and then scatter in the direction of the corresponding bin. This implies  that the travel direction of least resistance does not necessarily coincide with the direction of the detector. In a complex setup it is not clear  at the outset  which are the paths  of least resistance. 
As a testbed, the slab experiment allows us to make general statements regarding the interplay of albedo $A$, scattering order $n$, and the traveled path length $\Delta l$. In particular, we study the contribution of different $n$ to the total intensity. Solutions for the transmitted radiation through an illuminated slab have already been derived in the past  \citep[e.g.,][]{1975rmso.rept.....A}. However, we derive such solutions with a  novel approach (to our knowledge) for the one-dimensional slab in  Appendix \ref{sec:app:solution_one_dimensional_slab_problem} as a new perspective may have the potential to lead to new insights. 

Detected photon packages traveled through the one-dimensional slab, scattered in general several times, and eventually left the slab toward the detector. The probability $p^{(n)}$ that a photon package  undergoes $n$ scattering events before its detection is given by 
\begin{equation}
p^{(n)} = \frac{I^{(n)}}{I^{\rm tot}},
\end{equation}
where $I^{(n)}$ and $I^{\rm tot}$ are the $n$-th order contribution and the total intensity, which can be calculated using Eq. \eqref{eq:app:I_n} and Eq. \eqref{eq:app:I_tot}, respectively.

   \begin{figure}
   \centering
   \includegraphics{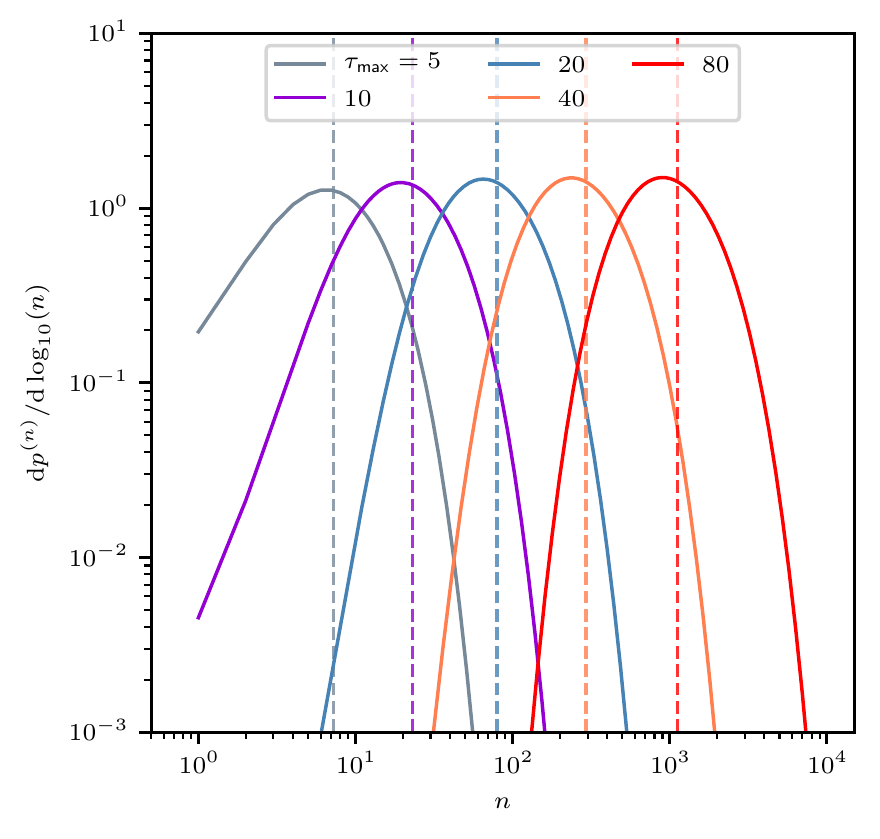}
   \caption{Detection probability for transmitted photon packages of scattering order $n$ through an one-dimensional slab. Plotted are the results for five different transverse optical depths $\taumax \in \left\{ 5,10,20,40,80\right\}$ under the assumptions $A=1$ and using isotropic scattering. Vertical dashed lines mark the mean scattering order of the corresponding optical depths. For details, see Sect. \ref{sec:app:transmission_trhough_one_dim_slab}.}
              \label{fig:oneD_slab_transmission}%
   \end{figure} 

Figure \ref{fig:oneD_slab_transmission} shows $p^{(n)}$ for the case where $A=1$ (i.e., for an absorption-free medium) for five different transverse optical depths $\taumax \in \left\{ 5,10,20,40,80\right\}$. As described in Sect. \ref{sec:app:transmission_trhough_one_dim_slab}, we used analytical solutions of the first few orders of Eq. \eqref{eq:app:I_n_integralform} as a measure for convergence, and truncated the sum when  an accuracy of $0.1\%$ was reached. In the case of $\taumax=80$, however, we truncated the sum at the order $k=10^9$, before reaching its convergence criterion for $n\leq4$. However, the displayed results show only solutions for significantly higher scattering orders, where the sum converges faster (see Appendix \ref{sec:app:transmission_trhough_one_dim_slab} for details). Figure \ref{fig:oneD_slab_transmission} shows a clear increase in the scattering order of highest probability $n^{\rm max}$ for increasing optical depths. The mean scattering order of detected photon packages, displayed as dashed vertical lines in the plot, was calculated using equation \eqref{eq:app:N_mean}, and shifts likewise with increasing $\taumax$ toward higher scattering orders and increasingly away from its corresponding $n^{\rm max}$ value. This shift is caused by the increasing contribution of higher scattering orders for increased optical depth (i.e., for $n > n^{\rm max}$ the slope of $p^{(n)}$ increases with increasing $\taumax$). Furthermore, we find that low scattering orders $n \leq n^{\rm max}$ are suppressed since they require a rapid displacement of the photon package toward the observer; in other words,  they exhibit a distinctly directed overall movement. However, high scattering orders $n\geq n^{\rm max}$ are also suppressed because these paths are confined to the volume of the slab (i.e., they experience an overall highly undirected displacement). This problem can be viewed as a pathfinding problem. Moreover, we find that for higher transverse optical depth $\taumax$, the range of contributing scattering orders increases strongly, which we refer to as the scattering order problem. In the case where $A<1$, the contribution of the $n$-th scattering order is linked to the contribution displayed in Fig. \ref{fig:oneD_slab_transmission} by multiplying $p^{(n)}$ with $A^n$ and rescaling. This results in a shift of $n^{\rm max}$ toward smaller $n$ values, meaning that the lower the albedo, the more quickly the photon package has to traverse the slab in order to have a significant contribution to its transmission spectrum. In the case where $A=0.5$, for instance, the mean number of interactions a detected photon undergoes for $\taumax=80$ is about $N^{\rm mean}\approx 30$, while for the same slab and $A=1$ the mean number of interactions is ${\sim}1119$. 

On the basis of these considerations it is possible to specify the advantages  and drawbacks of the composite-biasing technique. The main effect of this technique is to increase the average distance a photon package travels between two consecutive scattering events. This affects the path in two ways: first, by reducing the overall distance the photon package travels before being detected and second, by reducing the scattering order of the photon package. This can be seen by considering the path of a photon package that originates in the center of a dense homogeneous sphere, which travels and scatters multiple times inside it before leaving the sphere through its rim. The number of interactions $N_{\rm int}$ it undergoes before leaving the sphere  increases quadratically with the optical thickness $\tau_{\rm sph}$ of the sphere:  $N_{\rm int}\sim\tau_{\rm sph}^2$ \citep[e.g.,][]{2020A&A...635A.148K}. Increasing the traveled distance by some constant factor $f_x$ is equivalent to reducing the optical depth of the sphere by that factor (i.e.,  $N_{\rm int}\sim(\tau_{\rm sph}/f_x)^2$), thus it is effectively reducing the scattering order quadratically. The traveled distance of the photon package inside the dense sphere is on average given by $\Delta T = N_{\rm int} \Delta \tau$, where $\Delta \tau$ is the average traveled distance between two consecutive events of interaction and satisfies $\Delta \tau = f_x$. Therefore, the traveled distance satisfies $\Delta T\sim f_x^{-1}$, meaning that it is reduced by the factor $f_x$. As a result the composite-biasing technique generates paths of low scattering orders that reach the detector in a relatively short distance. Nonetheless, it is clear that a too high value of $\taumax$ and a high value of $A$ may lead to a situation in which the generated paths do not result in the desired range of scattering orders. Furthermore, it seems inevitable that very high $\taumax$ and $A$ values require the simulation of a broad range of observed scattering orders, as seen in Fig. \ref{fig:oneD_slab_transmission}. In order to sample photon package paths according to their importance to the total observed intensity it is thus mandatory to  generate high numbers of photon packages; in addition,  they should    exhibit a high number of scattering events. This, however,   adversely affects MCRT simulations that are equipped with the \sas method.


\section{Methods}
\label{sec:methods}

\subsection{Basic peel-off approach}
\label{sec:basic_peel_off}
The \sas method is expected to encounter issues at high optical depth and high albedo. In particular, Sect. \ref{sec:problems_high_optical_depth} demonstrates the necessity for covering a wide range of scattering orders, which is not feasible with the \sas method alone. There is, however, a promising method for solving this scattering order problem:  the peel-off method. This method is already used by many MCRT codes, for example, Mol3D \citep{Ober_2015} and POLARIS \citep{2016A&A...593A..87R}. It simply generates an additional photon package, the peel-off photon package, at every point of interaction of the original photon package. The peel-off photon package is then sent immediately (i.e., interaction-free) to the detector and carries a weight that is calculated according to the probability for the original photon package to be scattered in this direction. This weight is also reduced due to the extinction optical depth between the point of interaction and the observer. The weight of the peel-off photon package $w^{\rm po}_j$ which is detected by the $j$-th detector bin is given by
\begin{equation}
w^{\rm po}_j = \frac{\worg}{2 M} \exp\left( - \frac{\tau_{\rm tv}}{\mu_j}\right),
\end{equation}
where $w_{\rm org}$ is the weight carried by the original photon package and $\tau_{\rm tv}$ is the transverse optical depth between the point of interaction and the bottom boundary of the slab. Since every scattering event of a path is represented by an individual photon package, every simulated photon path increments the measured flux at the detector for a range of scattering orders. In particular, the range of simulated scattering orders ranges from zero to the total number of scattering events the photon package undergoes before leaving the slab. In later sections of this paper this peel-off method is referred to as the basic peel-off method. 
In Sect. \ref{sec:results_and_discussion} we present results for a comparison between the \sas method and the Split\,\&\,Peel-off\,\&\,Stretch (SPS) method. 

\subsection{Extended peel-off}
\label{sec:extended_peel_off}
According to the findings in Sect. \ref{sec:problems_high_optical_depth} it is clear that the basic peel-off method has its computational limits since the required number of simulated scattering events rapidly grows with $\taumax$ (i.e., with the volume of the slab), as can be seen in Fig. \ref{fig:oneD_slab_transmission}. Therefore, it is  expected that the \sps method fails to properly simulate arbitrarily high transverse optical depths and a more advanced method is needed. In order to compensate for the increased complexity of photon paths at higher $\taumax$, we propose a novel method, which we call the extended peel-off method. Later we combine this method with the \sas method and compare this Split\,\&\,extended\,Peel-off\,\&\,Stretch (SePS) method against the \sps method. 
Instead of generating a single peel-off photon package at a point of interaction, this method creates several peel-off photon packages that originate from an extended region. When the method is used, a homogeneous sphere needs to be defined with the point of interaction defining its center. Subsequently, the photon package is moved immediately to the rim of the sphere and the emission properties of the sphere are used to  emit this original photon package and to  generate peel-off photon packages at different points on the surface of the sphere. These features eventually enhance the performance of a MCRT simulation and improve the quality of its outcome (see Sect. \ref{sec:results_and_discussion}). For this method it is crucial to precalculate the emission properties of spheres of certain sizes since they are used for the photon package path determination and  for the flux calculation. This method aims to solve a few of the problems mentioned in Sect. \ref{sec:problems_high_optical_depth}.
First, emission spectra of spheres have to be precalculated only once and can then be used repeatedly during one simulation and even across different simulations. Therefore, if sufficient time is spent on the calculation of the spectra, all subsequent MCRT simulations will benefit. Second, photon packages move more quickly through the model space since they are repeatedly emitted from extended spheres, which saves computation time and allows the number of simulated photon packages to be increased. Third, since sphere spectra are composed of photon packages of different scattering orders, any peel-off photon package is also composed of contributions from multiple scattering orders. In particular, if the original photon package scatters for the $n$-th time, the generated peel-off photon package contains information of all orders $n_{\rm po}\in \left\{ n,n+1,n+2,...\right\}$, unlike the basic peel-off, which   generates a pure $n$-th order photon package. This is also the case for the original photon package that is emitted from the surface of the sphere. Moreover, Fig. \ref{fig:oneD_slab_transmission} shows an increasingly negligible contribution of low scattering orders with increasing optical depth. Therefore, simulations that are based on the basic peel-off method are required to simulate a high number of weakly contributing scattering orders before reaching the range of highly contributing scattering orders. Simulations that use the extended peel-off method, on the contrary, contain contributions of arbitrarily high scattering orders already from the first application of the method. As a result, the extended peel-off method results in an increased amount of computational time spent on actually relevant interaction events.

Before describing the steps we took for the precalculation of spheres and before showing their results, we briefly comment on the MRW and relate it to the extended peel-off method. The MRW is similar to the extended peel-off method insofar as it aims to solve the problem of high optical depths by using predetermined solutions of a sphere's spectrum and by quickly moving photon packages from the center to the rim of that sphere. In theory, its analytically obtained emission spectra could be used for the extended peel-off method and, consequently, the numerical MC based precalculations of these spectra could be avoided. However, there are two reasons why we decided against it. First, the MRW is limited in the type of scattering it is based on. Originally, it was developed for isotropic scattering \citep{1984JCoPh..54..508F} and then extended to anisotropic scattering \citep{1978wpsr.book.....I} where the scattering probability only depends on the angle between the incident and outgoing particle direction. A further extension to Mie-scattering \citep{1908AnP...330..377M}, however, is not possible, making a proper treatment of polarization impossible. Secondly, the emission properties of photon package off the surface of the sphere are not defined, rendering a precise extended peel-off method impossible too. In the following section, we describe how emission spectra of spheres can be precalculated and in Sect. \ref{sec:anisotropic_scattering_and_polarization} we discuss the inclusion of polarization to the extended peel-off method.

\subsubsection{Precalculation: Emission spectra}
\label{sec:precalc_emission_spectra}

   \begin{figure}
   \centering
   \includegraphics{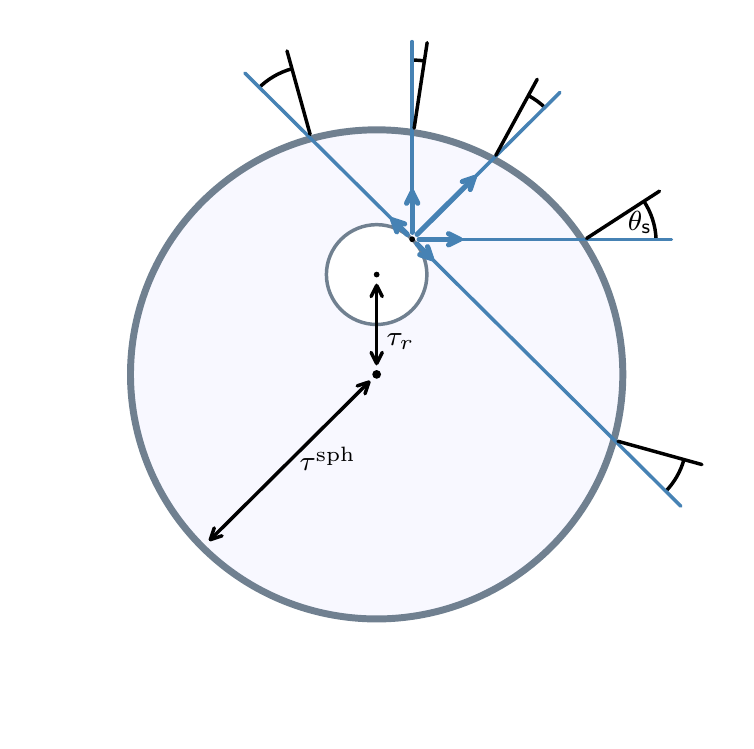}
   \caption{Sketch of a radiating sphere at radial position $\tau_r$ embedded in a larger sphere with radius $\tausph$. Radiation originating on the surface of the embedded sphere penetrates the surface of the larger sphere about the penetration angle $\theta_s$, and thus contributes to its emission spectrum. For details, see Sect. \ref{sec:precalc_emission_spectra}.}
              \label{fig:small_sphere_in_big_sphere}
   \end{figure}

We precalculate spectra of spheres with optical depth radii $\tausph\in \left\{ \tausph_k  \,\mid\, k \in \left\{1,2,3,4,5\right\} \wedge \tausph_k=10^{k/2}\right\}$. Transverse optical depths below $\tausph{=}\tausph_1$ are well represented by basic peel-off. The size of the largest precalculated sphere depends on the largest optical depth that can be encountered during a MCRT simulation, which is  $\taumax{=}300$ in this paper. We note,  on the one hand, that a higher number of precalculated spectra generally results in a more optimized simulation of photon package paths. On the other hand, it increases the time spent on precalculations. While a study of the optimal number and spacing of the values of $\tausph$ would lead to further improvement of the extended peel-off method, it is not part of this study. However, while the chosen number of precalculated spectra is rather low and thus unproblematic, the values chosen for $\tausph$ are  very useful, as will become clear in the course of this paper. In order to precalculate emission spectra that are based on isotropic scattering, we make use of the symmetry to reduce the amount of data that has to be stored. In Sect. \ref{sec:results_and_discussion} we discuss a more general case,  as well as the inclusion of polarization in the extended peel-off method. 

The precalculation setup is always composed of a sphere of certain radius $\tausph$ and a source of luminosity $1$ in its center. The particular unit of this luminosity is irrelevant, since we are solely interested in the relative luminosity that leaves the sphere. The luminosity is carried by a number of photon packages (we chose $10^8$) that are generated by the source, potentially scatter multiple times at various optical depth radii $\tau_r$, and then eventually leave the sphere across its surface at optical depth radius $\tausph$. The smallest sphere ($k=1$) is precalculated using the \sps method. Larger spheres ($k > 1$), on the other hand, use the \seps method by utilizing spectra of already precalculated smaller spheres. The luminosity that is emitted through the surface is stored in $100$ linearly sampled $\mu_s$ bins, where $\mu_s{\in}\left[0,1\right]$ is the cosine of the penetration angle $\theta_s$ (i.e., $\mu_s {=} \cos\left( \theta_s\right))$. Here, $\mu_s{=}1$ and $\mu_s{=}0$ are parallel and perpendicular, respectively, to the corresponding outwardly directed normal vector of the sphere. 

In order to apply the \seps method already during the precalculation of a larger sphere, a  precalculated smaller sphere has to be embedded inside it (see Fig. \ref{fig:small_sphere_in_big_sphere}). It is important to note that the \seps method in general can only be used if a photon package interacts and if the point of interaction and the closest border of the simulated homogeneous region are at a sufficiently large distance $\Delta \tau_{\rm min}$, in particular $\Delta \tau_{\rm min}\geq\tausph_1$. During the precalculation this condition becomes $\tausph-\tau_r \geq \tausph_1$. If the distance is not sufficient, we instead use the \sps method. 

Subsequent to a scattering event, there is a probability that the photon package will leave the sphere across its surface independent of the particular scattering direction. Hence, the peel-off luminosity can be calculated in every direction and assigned to the corresponding $\mu_s$-bin of the larger sphere in order to calculate the contribution to its emission spectrum. To do this we define the polar angle $\theta_\nu$ in the reference frame of photon package, such that $\theta_\nu=0$ and $\theta_\nu=\pi$ correspond to the direction of the smallest distance and of the largest distance to the surface of the larger sphere, respectively. In Fig. \ref{fig:small_sphere_in_big_sphere} $\theta_\nu=0$ and $\theta_\nu=\pi$  respectively correspond to the north and the south pole of the larger sphere. Additionally, we define $\nu = \cos\left( \theta_\nu\right)$ as the cosine of this angle. Next we describe how the peel-off emission is calculated for the case of the \sps method and the \seps method.

\paragraph{The \sps method:}

When the \sps method is applied during the precalculation the peel-off emission of the photon package in the direction $\nu$ contributes to the larger sphere's $\mu_s$-bin of direction
\begin{equation}
\mu_s(\nu) = \sqrt{1-\frac{\tau_r^2}{{\tausph}^2}\left( 1-\nu^2 \right)}.
\label{eq:mus_of_nu}
\end{equation}
This implies that a scattering event at radius $\tau_r$ results in a non-zero contribution for all $\mu_s\geq  \sqrt{1-\frac{\tau_r^2}{{\tausph}^2}}$. Thus, only scattering events close to the surface of the sphere contribute to low $\mu_s$-bins. The calculation of the full contribution to a particular $\mu_s$-bin can be obtained by solving the integral 
\begin{equation}
J(\nu)=\frac{1}{2} \int d\nu\, \exp\left( -x(\nu)\right),
\label{eq:integral_peel_peel_off_contribution}
\end{equation}
where $x(\nu)=-\tau_r \nu + \sqrt{{\tausph}^2-\tau_r^2\left( 1 - \nu^2\right)}$ is the distance to the surface in direction $\nu$. The portion of $\worg$, which crosses the surface of the sphere in a specific $\mu_s$-bin is obtained by choosing the $\nu$-limits in the integral that correspond to the $\mu_s$-bin by using Eq. \eqref{eq:mus_of_nu}. An integral of Eq. \eqref{eq:integral_peel_peel_off_contribution} is given by
 \begin{equation}
   J(\nu) = 
  \begin{cases} 
      \frac{1}{2}\exp(-\tausph)\nu &\text{if }\tau_r=0,\\
      \frac{1}{4\tau_r} \Bigl[ \exp(-x(\nu)) + \left({\tausph}^2-\tau_r^2 \right)\boldsymbol{\cdot}\cdots &\\
    \hfill \cdots\boldsymbol{\cdot} \left( \frac{\exp(-x(\nu))}{x(\nu)} + {\rm Ei}(-x(\nu))  \right) \Bigr]&\text{if }0<\tau_r<\tausph,\\
      \frac{1}{2}\nu&\text{if }\tau_r=\tausph \wedge \nu>0,\\
      \frac{1}{4\tausph} \left[\exp(2\tausph \nu) - 1 \right] &\text{if }\tau_r=\tausph \wedge \nu\leq0,\\
   \end{cases}
   \label{eq:precalc_basic_peel_off}
 \end{equation}
where ${\rm Ei}(x)$ is the exponential integral defined as
\begin{equation}
{\rm Ei}(x) = - \int_{-x}^\infty ds\, \frac{\exp(-s)}{s}.
\end{equation} 

\paragraph{The \seps method:}

If the distance between a scattering position and the surface of the sphere exceeds $\tausph_1$, a smaller  precalculated sphere can be placed inside. This scenario is depicted in Fig. \ref{fig:small_sphere_in_big_sphere}. Hereafter the spectrum of the small sphere can be used to determine its transmission through the surface of the larger sphere it is embedded in.

   \begin{figure}
   \centering
   \includegraphics{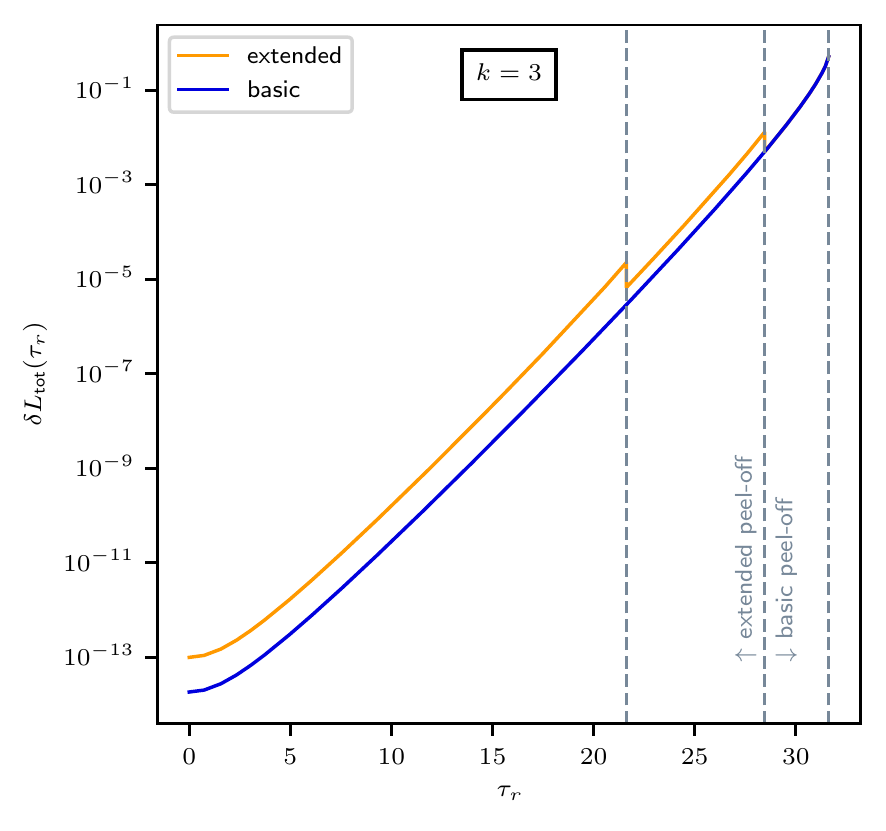}
   \caption{Total transmitted peel-off contribution $\delta L_{\rm tot}$ of radiating embedded spheres or points, inside a larger sphere of size $\tausph_3=10^{1.5}$. The orange and the blue curve show the contribution when using the extended and the basic peel-off method, respectively. The three gray vertical lines show the transition radii $\tau_r$, where a differently sized sphere can be embedded. In particular, the left gray line gives the transition from using a sphere of size $\tausph_2$ to a sphere of size $\tausph_1$. The middle line separates the $\tausph$ region from the region where the basic peel-off is used. Close to the rim of the  sphere (i.e., to the right of the last gray line for radii $\tau_r \lesssim \tausph_3$) none of the precalculated spheres fit and the basic peel-off method is applied.}
              \label{fig:contribution_subsphere_position}
   \end{figure} 

The transmission spectrum of such an embedded sphere strongly depends on $\tau_r$.  However, its calculation involves the solution of a computationally expensive multi-dimensional integral since radiation is emitted from every point of the surface of the embedded spheres and in every direction of its corresponding hemisphere. As displayed in Fig. \ref{fig:small_sphere_in_big_sphere}, the transmitted luminosity must also  be sampled in the larger spheres $\mu_s$-bins. Therefore, before precalculating the spectrum of a larger sphere, we first precalculate the contribution of all embedded spheres to its emission spectrum for a set of selected $\tau_r$ values. In this way  we can quickly update the emission spectrum of a larger sphere during its precalculation by accessing and interpolating these   previously precalculated transmission spectra. 

To ensure a proper sampling of precalculated $\tau_r$ values, we make use of the fact that any embedded sphere is always chosen to be as large as the geometrical limitations allow. Therefore, the deeper the scattering event of the original photon package occurs, the larger the embedded sphere is chosen. As a consequence, the interval for radial positions $\tau_r\in\left[ 0, \tausph \right]$ can be divided into disjoint intervals, each corresponding to the size of an embedded sphere. In particular, larger values of $\tau_r$ correspond to smaller embedded spheres as the distance to the surface of the larger sphere is smaller. In the interval $\left( \tausph-\tausph_1,\tausph\right]$, however, no sphere can be embedded, and instead the transmission spectrum is given by the basic peel-off by Eq. \eqref{eq:precalc_basic_peel_off}. As pointed out earlier, transmission spectra of embedded spheres only have to be evaluated within their corresponding $\tau_r$-intervals. First we determine the contribution to the emission spectrum for an embedded sphere at the infimum and the supremum of its corresponding $\tau_r$-interval. Considering  that different $\mu_s$-bins may be accessible depending on the position of the embedded sphere, we then additionally calculate the contribution at those points where the number of accessible bins changes. In other words, the closer the embedded sphere is placed to the surface of the larger sphere, the more $\mu_s$-bins receive flux from the radiating embedded sphere. Using this particular sampling of evaluated transmission spectra ensures that during their interpolation $\mu_s$-bins receive a contribution only if they actually have a non-zero contribution. The number of evaluated radial positions for different sphere sizes $\tausph$ therefore differs, and in particular the number of evaluated positions inside spheres of size $\tausph_2$, $\tausph_3$, $\tausph_4$, and $\tausph_5$ are 97, 161,199, and 221, respectively.


In Fig. \ref{fig:contribution_subsphere_position}, we show as an example for $\tausph=\tausph_3$ the results for the total contribution $\delta L_{\rm tot}$ to the emission spectrum. For this calculation we used $10^8$ photon packages. We note  that the transmission spectrum is sampled with respect to the spheres' $\mu_s$-bins; however,  $\delta L_{\rm tot}$  is the sum of all contributions. The blue curve shows the total luminosity that is transmitted due to basic peel-off for a neutral photon package that  has interacted at the radial position $\tau_r$. In general, $\delta L_{\rm tot}$ increases with increasing $\tau_r$ and reaches its maximum of $\delta L_{\rm tot}{\gtrsim} 0.5$ at $\tau_r=\tausph_3$. Compared to the basic peel-off, the extended peel-off, which is shown as the orange curve, generally reaches higher values of $\delta L_{\rm tot}$. This is a consequence of the fact that it takes into account higher scattering orders than the basic peel-off. Moreover, the curve shows discontinuities which result from the use of a discrete set of precalculated spheres. Deeper inside (i.e., for lower values of $\tau_r$) larger precalculated spheres can be embedded. The first vertical gray line represents the highest value of $\tau_r$ at which spheres of size $\tausph_2$ can be embedded. The second vertical gray line marks the highest radius for the embedment of a sphere of size $\tausph_1$. Past this radius none of the precalculated spheres fit inside and the basic peel-off method is used.

The calculation of the transmitted spectrum and the corresponding total transmitted luminosity $\delta L_{\rm tot}$ is performed by the use of MC integration. That means, non-interacting photon packages are sent from the surface of the embedded sphere in a probabilistically determined direction and cross the surface of the larger sphere with a certain penetration angle. The weight these photon packages carry after traversing the required optical depth to reach the surface, is then added to the corresponding $\mu_s$-bin. In this case, the launching position on the surface of the embedded sphere is chosen isotropically and the direction of emission is chosen randomly according to the emission properties of the embedded sphere. The resulting contribution to the emission spectrum is eventually normalized, such that it corresponds to an originally neutral photon package. It is important to note that these results are only used for the precalculation itself, for which they are specifically performed. During a realistic MCRT simulation, only the emission spectra of the precalculated spheres will be used. 

\paragraph{Precalculated sphere spectra:}

\begin{figure}
   \centering
   \includegraphics{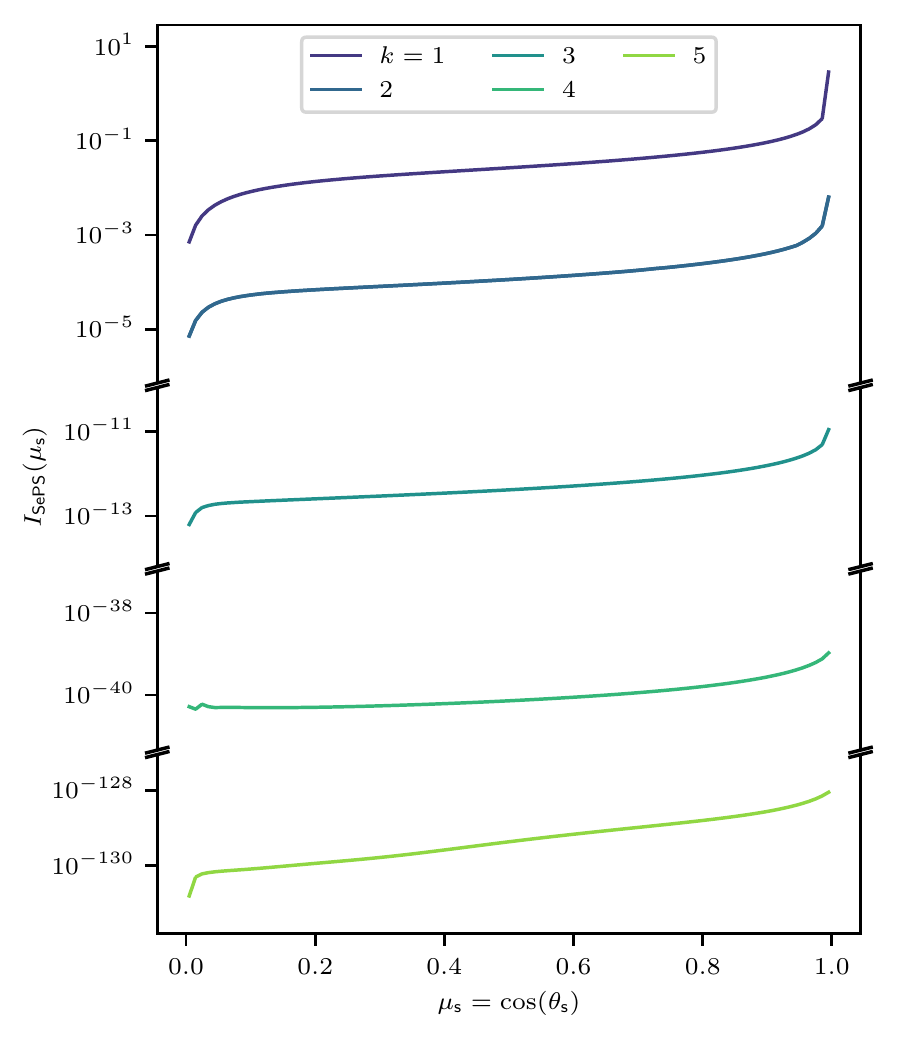}
   \caption{Precalculated emission spectra for spheres of optical depth radius $\tausph_k$ for $k\in\left\{1,2,3,4,5\right\}$. The internal source of luminosity is neutral and isotropically radiating. It is placed in the center of the sphere and propagates through the sphere until it leaves it through its rim. The intensity $I_{\rm SePS}$ leaves the surface of the sphere and depends on the cosine of the  penetration angle $\mu_s=\cos(\theta_s)$. The intensity is highest in the direction radially outward and overall decreases toward a smaller $\mu_s$.}
              \label{fig:reference_spectrum}
   \end{figure}

\begin{figure}
   \centering
   \includegraphics{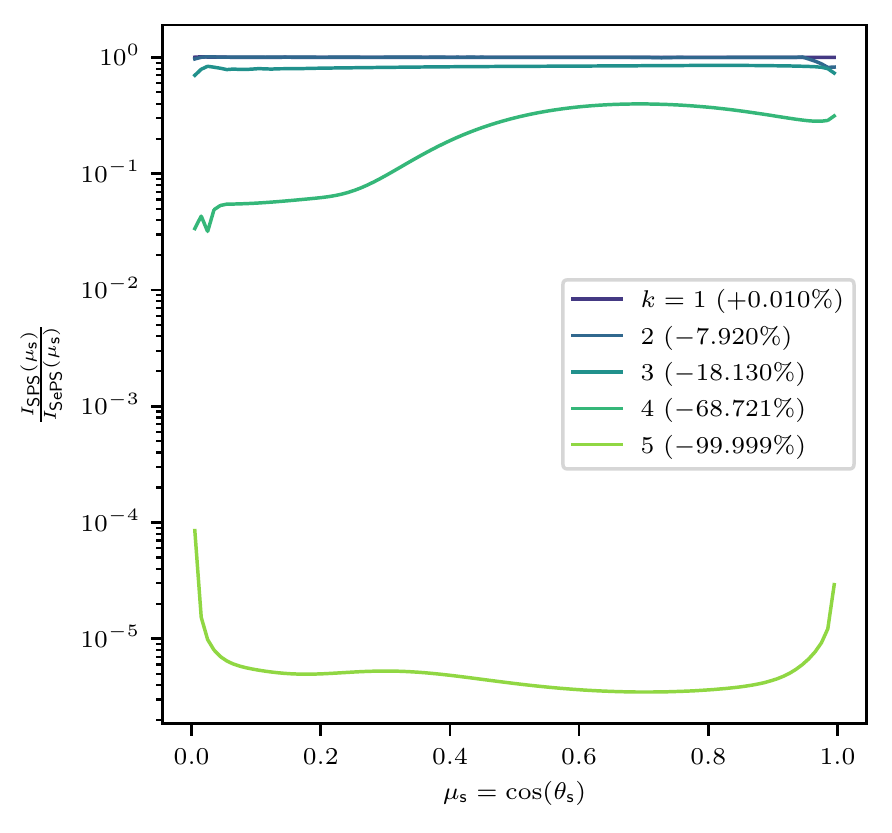}
   \caption{Ratio of emission spectra $I_{\rm SePS}$ to $I_{\rm SPS}$, which use the \seps method and the \sps method, respectively, during the precalculation. The \sps method reaches increasingly lower intensities for increasing sphere sizes $\tausph_k$ (i.e., for higher values of $k$). Shown in brackets  is the total loss of emitted luminosity when using the \sps method compared to the \seps method. For details, see Sect. \ref{sec:precalc_emission_spectra}.}
              \label{fig:spect_comparison_nPO}
   \end{figure}

Figure \ref{fig:reference_spectrum} finally shows the obtained emission spectra of five precalculated spheres of optical depth radius $\tausph_k$ for $k\in\left\{1,2,3,4,5\right\}$. The highest intensity of any sphere is emitted in the radial direction ($\mu_s=1$) and decreases toward lower $\mu_s$. For this calculation we used the \seps method whenever possible. It is crucial to use a method like this to precalculate the spectra, since the use of the basic peel-off strongly underestimates the resulting emission spectra, which can be seen in Fig. \ref{fig:spect_comparison_nPO}. It shows the ratio of the emitted intensity for different sphere sizes when using the \seps method  to the \sps method during the precalculation. Small sphere sizes (i.e., small values of $k$) do not show a strong deviation between the two methods. However, the relative differences increase for increasing sphere sizes and the intensity is clearly underestimated for the full range of $\mu_s$ values. The plot also shows in brackets the total estimated luminosity a sphere loses when utilizing the \sps method rather than the \seps method, which for $k=4$ (i.e., $\tausph_4=100$) already reaches ${\sim}68\%$ and for $k=5$ (i.e., $\tausph_5\approx310$) even ${>}99\%$.

\subsubsection{Precalculation: Projection onto detector}

One crucial step in a realistic MCRT simulation is the determination of flux. To this end, a detector that measures incoming flux needs to be defined as an object of certain geometrical size and orientation. Most MCRT codes, in particular the previously mentioned codes, usually implement the detector as a geometrically confined plane object at a great distance with an orientation that can be characterized by a constant normal vector. The infinite slab setup introduced in Sect. \ref{sec:slab_numerical_implementation}, however, has a curved detector with non-constant normal vector. In this regard, the detector of this specific problem is comparably complex. 

First we describe a method for propagating extended peel-off emission from an embedded sphere to the detector. To this end, the surface of the sphere is divided into patches whose radiation is projected onto the detector. Subsequently, this method is applied to the detector of the infinite slab setup. The application to a flat detector at great distance is  discussed toward the end of this chapter and is merely a simplified variant of the presented method. 

Whenever the original photon package interacts and its distance to any boundary of the slab is sufficient, the point of interaction defines the center of a precalculated sphere of size $\tausph_k$ for some value $k$ that is chosen to be as large as possible. This sphere is then embedded in the slab, and every point of its surface radiates in its corresponding hemisphere. During the application of the peel-off method, radiation whose direction satisfies $\mu>0$ in the frame of the detector stays interaction free and is thus eventually observed by the detector. Since the sphere in embedded in a region of non-zero density, different points on its surface may result in different detected peel-off intensities, and the particular emission properties may also have angular dependencies. In order to account for these features, we divide the surface of the sphere into $N_p$ patches. In order to properly represent the emission of a sphere by patches, they are required to be  limited by a maximum deviation of the normal vectors and to be  limited by a maximum extent in optical depth units. The first criterion is met if any two points $\vec{p}_1$ and $\vec{p}_2$ in a patch and their respective normal vectors $\vec{n}_1$ and $\vec{n}_2$ fulfill $\vec{n}_1\cdot\vec{n}_2 \leq C_n$ for some chosen constant $C_n>0$. The second criterion can be satisfied by imposing the condition $\abs{\vec{p}_1-\vec{p}_2}\leq C_p$ for another chosen constant $C_p>0$. For increasingly large spheres, the first directional criterion is met easily as the curvature of the surface decreases with its radius. Therefore, the spatial extent of the patches becomes the dominant restriction. However, due to the azimuthal symmetry of the infinite slab setup, we apply a simpler criterion and choose patches such that any two surface points $\tilde{\vec{p}}_1$ and $\tilde{\vec{p}}_2$ with equal azimuthal position within a patch fulfill
\begin{equation}
\abs{\tilde{\vec{p}}_1-\tilde{\vec{p}}_2} \leq \Delta \tau_{max}^p,
\label{eq:patches}
\end{equation}
where we choose $ \Delta \tau_{max}^p = 1$. The condition Eq. \ref{eq:patches} results in circular patches at the poles and ring-like patches between these poles, which are symmetric with respect to the transverse direction of the slab. These resulting patches have a constant extent in their polar angles. In the case of the sphere size $\tausph_1$, $\tausph_2$, $\tausph_3$, $\tausph_4$, and $\tausph_5$ the number of patches is $N_p=12$, $20$, $36$, $64$, and $112$, respectively. 

\begin{figure}
   \centering
   \includegraphics{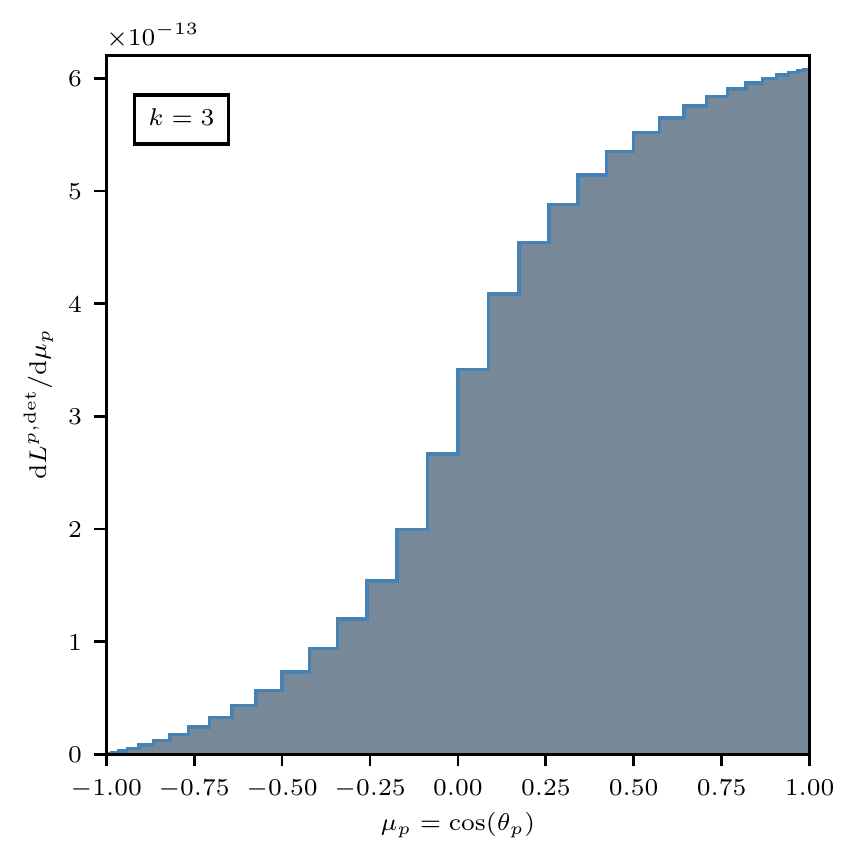}
   \caption{Emitted luminosity per patch size ${\rm d} L^{p,{\rm det}}/{\rm d}\mu_p$ as a function of the position $\mu_p=\cos(\theta_p)$ on the surface of the sphere. The area beneath the curve corresponds to the total emitted luminosity of the sphere. Values of $\mu_p$ increase when nearing a face-on view of the detector; resulting in a rising extended peel-off contribution.}
              \label{fig:L_patch_detector}
   \end{figure}

It is possible to precalculate the luminosity $L_j^p$ that is emitted from any patch $p$ in the direction of the $j$-th detector bin. Every patch then emits this luminosity from its center, which is defined by the average $\mu$ value of the patch, and sends it toward the detector bins. It is convenient to use luminosity as a measure for the extended peel-off method rather than intensities since the weight of a photon package is associated with its carried luminosity. In a MCRT simulation this peel-off luminosity is additionally attenuated by the material on the path between the patch and the detector. The luminosity $L_j^{p,{\rm det}}$ that the $j$-th detector bin measures due to the peel-off emission of the patch $p$, is given by
\begin{equation}
L_j^{p,{\rm det}} = L_j^p \exp\left(-\frac{\tau_p}{\mu_j}\right),
\end{equation}
where $\tau_p$ is the transverse optical depth between the launching point on the patch and the boundary of the slab. Figure \ref{fig:L_patch_detector} shows the example of  a sphere of size $\tausph_3$ and its total patch luminosity $L^{p,{\rm det}}=\sum_j L_j^p$ per patch size, which is a measure for the emitted flux of the sphere that is observed by the detector. It shows a strong increase in emitted flux toward higher values of $\mu_p$, which is the $\mu$ value that describes the location on the sphere. In other words, patches that face the detector contribute more strongly to the detected intensity. However, this plot also shows that all patches contribute and how the $\mu_p$ range of different patches varies as a consequence of the definition in Eq. \ref{eq:patches}. We note that the area below the curve equals the total observed luminosity $L^{\rm det}=\sum_{j,p} L_j^p$.

In order to precalculate the projected luminosity of any patch onto the detector we use a MC method in which we generate $10^7$ photon packages on the surface of a precalculated sphere. The launching position is chosen randomly according to an isotropic distribution. Then the direction of a photon package is determined randomly according to the spectrum of the sphere, which has an angular dependence as can be seen in Fig. \ref{fig:reference_spectrum}. Next, the $\mu$ value of that photon package is determined in the frame of the detector. If it is directed toward a detector bin, the luminosity that is carried by the photon package is added to the projected luminosity of the sphere. Due to the symmetry, half of these photon packages would not contribute to the projected intensity, since their traverse direction is opposite to that of the detector. To increase the performace and get a better estimate for the projected spectra even of patches that barely radiate toward the detector, we used a biasing technique and only chose photon package directions that are pointing toward the detector. To compensate for this the weight of these photon packages is reduced accordingly. Contrary to the precalculated transmission spectra described in Sect. \ref{sec:precalc_emission_spectra} that were solely used during the precalculation, the projected peel-off luminosities that we precalculate and that are described in this section are repeatedly used during the actual MCRT simulation as we describe later in Sect. \ref{sec:results_and_discussion}.

It should be noted that this elaborate method for calculating the projected luminosity is a consequence of the curvature of the detector. In the previously mentioned case of a flat detector at a large distance that can be described with a unique normal vector, the projected luminosity can be estimated with the flux in exactly that unique direction of the detector. Therefore, this case does not require a MC integration method. 

In addition to the described extended peel-off method, we test a simplified version of the extended peel-off method, later referred to as the simple \seps method. In this variant, the whole sphere is represented by a single peel-off photon package, rather than $N_p$ peel-off photon packages, and the luminosity leaving the sphere toward the $j$-th detector bin ($L_j$) is obtained by assuming a great distance to the detector. Hence, $L_j$ can be written as
\begin{equation}
L_j =  \frac{L_{\rm sph}^{\rm tot}}{2 M},
\end{equation}
where $L_{\rm sph}^{\rm tot}$ is the total luminosity emitted by the sphere. The peel-off photon package is then emitted from the point that is geometrically closest to and in the direction of the detector (i.e., the point satisfying $\mu_p=\mu_i$). 

In Sect. \ref{sec:results_and_discussion} we present results for a comparison between the \sas method, the \sps method, the \seps method, and the simple \seps method.

\section{Results and discussion}
\label{sec:results_and_discussion}

In this section the previously described \sps method (Sect. \ref{sec:basic_peel_off}) as well as the \seps and the simple \seps method (Sect. \ref{sec:extended_peel_off}) are applied to the infinite plane-parallel slab problem (see Sect. \ref{sec:setup_and_problem_analysis}). Furthermore, we present a brief discussion of the role of the albedo and of the extension of the \seps method with regard to anisotropic scattering and polarization. For  clarity, a short overview with descriptions of the different applied MCRT methods is provided in Table \ref{tab:overview_mcrt_methods}.

\subsection{Role of peel-off in MCRT simulations}
\label{sec:role_of_peel_off_on_mcrt}
\begin{figure}
   \centering
   \includegraphics{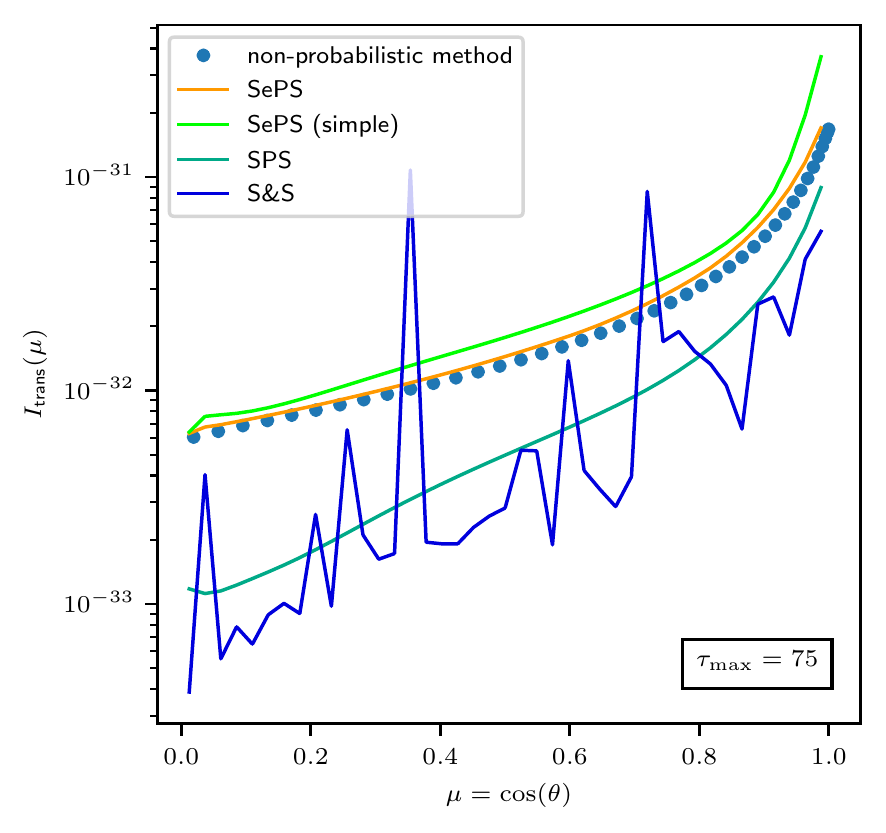}
   \caption{Results for the infinite plane-parallel slab problem with transverse optical depth $\taumax=75$. Different advanced MCRT methods (solid lines) are compared to results from a non-probabilistic numerical method (dotted line).}
              \label{fig:SePS_vs_SPS_vs_CB}
   \end{figure}

In Fig. \ref{fig:SePS_vs_SPS_vs_CB} the results of simulations of the infinite slab with $\taumax=75$ are shown. The dotted curve is the result of a non-probabilistic numerical calculation (see Sect. \ref{sec:non-probabilistic_method}) and is used for comparison with the results of different MC based methods. A total of  $10^9$ photon packages are used in all MCRT simulations. The plot clearly shows that the three simulations based on  the peel-off method    result in overall smoother results for the transmitted flux as opposed to the \sas method, which results in a  noisier transmission spectrum. This is a consequence of the fact that  during every scattering event of an original photon package a peel-off based simulation sends individual peel-off photon packages toward every detector bin, whereas the \sas method solely increments the detected flux of a single bin per original photon package. The \sps method shows an overall better performance than the \sas method in the sense that its overall detected luminosity and also the shape of the result more closely resemble the non-probabilistic solution. Nonetheless, both methods deviate by about one order of magnitude from the non-probabilistic solution at  low $\mu$ values. For increasing $\mu$, these relative differences on average decrease. Our results indicate that using the \sps method results in an improved estimate of the transmitted flux. Nonetheless, the overall underestimation of flux shows that the \sps method does not solve the problem of high optical depths. 

The \seps method and the simple \seps method achieved a significantly better result. The \seps method in particular shows a low relative error for all directions $\mu$, and is thus a suitable method for transverse optical depths of up to ${\sim}75$. The simple \seps method, however,   slightly overestimates the resulting flux. Toward higher values of $\mu$ this effect is anticipated,  first because the complete emission of the sphere is represented by a single photon package when using the simple \seps method, and second because this peel-off photon package is launched at a position close to the pole of the sphere that is facing the detector. Therefore, radiation that would otherwise originate from points of the sphere that are more deeply embedded in the slab are now experiencing a lower attenuation, resulting in a higher flux level. This effect can be expected to intensify for larger spheres and additionally lead to an underestimation of flux at lower $\mu$ values for similar reasons. For this reason the simple \seps method fails to produce proper flux estimates at higher transverse optical depths. 

Whether a method eventually succeeds or fails in the context of high optical depth MCRT simulations depends on the computation time it requires and the accuracy it achieves. The performance gain due to the \seps method then becomes apparent when considering the computation time required. All the presented simulations were performed serially (i.e., with a single core\footnote{Processor: Intel Xeon Gold 6148}) using \textsc{Python}. The calculation for the \seps method took ${\sim}2$\,days compared to the ${\sim}52$\,days of equivalent serial computation time reported by \cite{2018ApJ...861...80C}, who were using the \sas method with $10^{13}$ photon packages. The latter simulation could be expected to require a far greater number of photon packages to reach the same error as our \seps based simulation, which in addition used $10^4$ times fewer photon packages. As a result, the \seps method saved ${>}95\%$ of the simulation time and achieved a significantly better result. Even though a direct comparison between the two simulations is difficult, as expected, the \seps method clearly outperforms the \sas method at medium to high optical depths and proves to be the preferred method for this type of MCRT simulations.

\begin{figure*}
   \centering
   \includegraphics{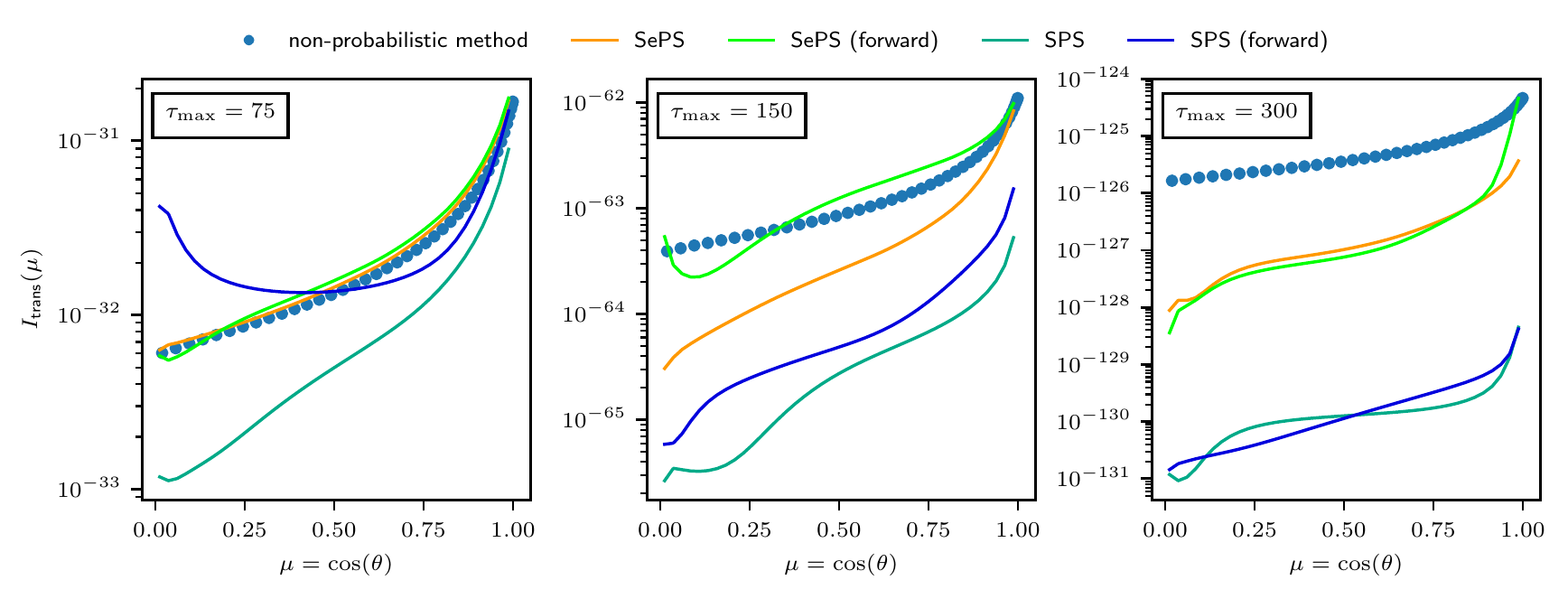}
   \caption{Results for the infinite plane-parallel slab problem for transverse optical depths $\taumax\in\left\{ 75,150,300\right\}$. Different advanced MCRT methods (solid lines) are compared to results from a non-probabilistic numerical method (dotted lines). The forward label indicates a variant of the \seps method that differs only in its application of the Stretch method; it is applied if and only if the photon package scatters toward the detector.}
              \label{fig:SePS_and_SPS_with_2dir_and_1dir}
   \end{figure*}

To study the feasibility of performing MCRT simulations at higher optical depths, Fig. \ref{fig:SePS_and_SPS_with_2dir_and_1dir} shows the results of four different MCRT simulations for three transverse optical depths $\taumax\in\left\{ 75,150,300\right\}$. In every simulation $10^9$ photon packages were used, which is the same number as   used previously for the results in Fig. \ref{fig:SePS_vs_SPS_vs_CB}. In addition to the \sps and \seps method, we also consider the forward \seps and the forward \sps method. These two methods differ from the \seps method and the \sps method, respectively, only in that the Stretch method is applied if and only if the direction of the photon package after scattering points toward the detector (i.e., in forward direction). As a consequence, methods using the forward label result in detected photon packages experiencing  on average a reduced number of scattering events and having traveled a shorter distance before leaving the slab. For all three values of $\taumax$ the total detected luminosities of methods that use the forward label exceed their respective bi-directional methods. Already at an optical depth of $\taumax=150$ the \seps method struggles to produce the correct result. The forward \seps method, however, is closer to the non-probabilistic solution.

The \seps method particularly underestimates the flux toward smaller $\mu$ values, which mainly originated from photon packages that interacted close to the bottom boundary of the slab. A lack of flux may therefore indicate a lack of representative scattering events close to this boundary. A potential solution to this problem may involve the introduction of a lower limit of scattering events that a photon package has to undergo before it can leave the slab. This is closely related to the problem of simulating the path of photon packages to sufficiently high scattering orders. It may very well be that too many photon packages are leaving the slab too early while still carrying a significant weight, both during the infinite slab simulation or already during the precalculation of sphere spectra. The fact that photon packages are allowed to leave the model space may at this point become a limiting factor as it systematically leads to an underestimation of flux. However, this raises the question of the order of the \mbox{minimum} number of considered scattering events, which may be the topic of future studies. Additionally, the results shown in Fig. \ref{fig:oneD_slab_transmission} also imply an upper limit for relevant scattering orders; this means that photon packages that exceed this number of scatter orders may safely be removed from the model space without significantly changing the final result for the estimated fluxes. Another potential source for the underestimation of flux may reside in spectra that have not been precalculated with sufficient accuracy. To ensure a sufficiently high accuracy of the precalculated spectra, and thus the possibility to control this additional source of error, it is highly recommended to use statistical tests for convergence, such as those described by \cite{2018ApJ...861...80C}. In this study, however, the number of simulated photon packages is fixed. 

At $\taumax=300$, none of the considered methods is capable of reproducing the non-probabilistic solution. These three plots also suggest that for increasing transverse optical depth the resulting flux level obtained by the \sps method decreases more strongly than the flux level of the \seps method,  meaning that it becomes increasingly relevant to use an advanced method like the \seps method at higher optical depths. Nonetheless, at $\taumax=300$ the estimated flux based on the \seps method is orders of magnitude too low. Only at $\mu{\lesssim}1$ does the forward \seps method predict fluxes that are compatible with non-probabilistic solutions for all three considered $\taumax{\leq}300$.

Therefore, we conclude that despite the success of the \seps method at an optical depth of $\taumax=75$, it is nonetheless not fully capable of solving the problem of MCRT simulations at high optical depths and it is questionable whether this problem can be handled by purely MC based methods. However, we find that the \sps method and the \seps method both benefit MCRT simulations in the regime of medium to high optical depths where fluxes may be severely  underestimated as they significantly alleviate this problem.  The main reason for their success lies in the coverage of a wide range of scattering orders, which is especially effective in the case of the \seps method,   designed to reach high scattering orders already during early stages of the life cycle of every photon package.

\paragraph{The role of the albedo:} \label{sec:the_role_of_albedo}

As described in Sect. \ref{sec:problems_high_optical_depth}, the albedo plays a decisive role in MCRT simulations at high optical depths. Similar to the effect of an increased optical depth $\taumax$, which can be seen in Fig. \ref{fig:oneD_slab_transmission}, an increased value of $A$ shifts the scattering order with the highest contribution toward higher values. Simultaneously, the range of highly contributing scattering orders also increases. However, the probability of simulating a path that contributes significantly decreases with its scattering order. Consequently, keeping the number of simulated photon packages fixed, it becomes less likely to generate a sufficient number of paths that are contributing significantly to the measured flux. In conclusion, higher values of the albedo $A$ require a greater number of simulated photon packages for a proper flux determination. Therefore, increasing the albedo results in a lower value of $\taumax$ that can be simulated with acceptable effort; in turn, a lower albedo results in a higher limit for $\taumax$. This can be seen in Fig. \ref{fig:SePS_for_largerA} and Fig. \ref{fig:SePS_for_smallerA} in the Appendix, which  show examples of the results of MCRT simulations for transmission spectra through a plane-parallel infinite slab for an increased albedo of $A=0.9$ and a decreased albedo of $A=0.05$, respectively. Figure \ref{fig:SePS_for_largerA} illustrates that the limit for $\taumax$ even drops below $\taumax=75$ when increasing the albedo to $A=0.9$, as the determined flux is strongly underestimated. This can be explained by the fact that for $A=0.5$ the limit for $\taumax$ lies between $\taumax=75$ and $150$. Increasing the albedo significantly reduces this limit. However, Fig. \ref{fig:SePS_for_smallerA} shows the results of MCRT simulations for a reduced value of the albedo of $A=0.05$, which is a typical value for dust in the micrometer wavelength range. It shows the results for transmission spectra through slabs of four different optical depths $\taumax\in\left\{ 75,150,300, 600\right\}$. For this low value of $A$ the $\taumax$ limit increases strongly. In particular, only at an optical depth of $\taumax=600$ do the results of the MCRT simulation using the \seps method start to show slight differences from the non-probabilistic solution. In conclusion, we find that a high albedo of $A=0.9$ has its limit at $\taumax<75$, while a low albedo of $A=0.05$ allows simulations up to $\taumax \gtrsim 600$. We conclude that increasing the albedo shifts the onset of the scattering order problem toward lower values of $\taumax$; in other words,  it reduces the $\taumax$ limit, which is in agreement with the findings in Sect. \ref{sec:problems_high_optical_depth}.

\subsection{Extension to the \seps method: Anisotropic scattering and polarization}
\label{sec:anisotropic_scattering_and_polarization}
In this study, we introduced the extended peel-off method and specifically applied it to the infinite plane-parallel slab problem studied by \cite{2018ApJ...861...80C}, which is based on purely isotropic scattering. In the following we discuss anisotropic scattering and polarization and how it can be integrated in the extended peel-off method. 

Henyey-Greenstein (HG) scattering \citep{1941ApJ....93...70H} is a widely used one-parameter model for anisotropic scattering. By integrating it into the extended peel-off method, as described in Sect. \ref{sec:extended_peel_off}, the resulting spectra for precalculated spheres become in general dependent on the initial direction of the photon package. Depending on the type of the detector, there are a few possibilities for successfully implementing this dependency. In the case of a curved detector, as is used in the infinite slab problem, patches were defined on the surface of precalculated spheres, which were oriented in the coordinate system of the detector. To implement HG scattering, the emission spectra of these patches have to be calculated separately for the full sampled range of initial directions of the photon package. In particular, the initial $\mu=\cos(\theta)$ value of the photon package needs to be sampled. While in the isotropic case the emission spectra of any point of the surface depend solely on the parameter $\mu_s=\cos(\theta_s)$, the anisotropic case leads to a dependence on a second angle, $\phi_s$, which describes the direction in the tangent plane of that point. Assuming a sampling of the angular parameters $\theta$, $\theta_s$, and $\phi_s$ in steps of $1^\circ$ as well as single precision ($4$ bytes per entry), the required memory per wavelength and patch is on the order of only ${\sim}200$ megabytes. In the case of a plane detector at a distant location, these patches can instead be defined in the coordinate system of the scattered photon package. This removes the necessity of sampling the initial direction of the photon package and, as a consequence, significantly reduces the memory consumption to ${\sim}1$ megabyte per wavelength and patch. The implementation for both types of detectors is therefore readily feasible. 

The inclusion of polarization to the extended peel-off method is possible as well, but leads to higher memory demands. Polarization is an important quantity that can be simulated in MCRT simulations. A proper treatment of polarization is often achieved by the use of Mie-scattering, whose outcome of scattering events depends on and affects the polarization state of a photon. It opens up many possibilities for in-depth studies of astrophysical systems, providing complementary insights to pure fluxes. 
The \sps method can utilize polarization in a straightforward manner and already constitutes a crucial part of many MCRT codes like Mol3D \citep{Ober_2015} and POLARIS \citep{2016A&A...593A..87R}. Compared to the inclusion of HG scattering, precalculated sphere spectra additionally depend on the initial polarization state of the scattered photon package. Precalculated spectra are meant to be calculated for neutral photon packages, where the weight for the initial photon package is associated with $I_{\rm in}$, which is the first component of the Stokes vector. The remaining Stokes components $Q_{\rm in}$, $U_{\rm in}$, and $V_{\rm in}$, however, need to be sampled, and they increase the memory consumption of precalculated spectra. Since memory demand may quickly pose a problem, the choice of sampled $Q_{\rm in}$, $U_{\rm in}$, and $V_{\rm in}$ is critical. In general, these parameters are limited by the inequality 
\begin{equation}
I_{\rm in}^2 \leq Q_{\rm in}^2+U_{\rm in}^2+V_{\rm in}^2,
\end{equation}
and  sampling   them may be realized for instance by using a grid of cartesian or spherical coordinates.  This means that  whenever the extended peel-off method is applied the three-dimensional bin corresponding to the Stokes vector of the scattered photon package is determined and its precalculated spectrum subsequently used. Assuming a cartesian sampling of these parameters in steps of $10\%$ thus results in another factor of ${\sim}4000$ for the memory consumption. Moreover, photon packages that cross the rim of the sphere are also polarized, resulting in another factor of $4$ in the memory consumption to store all components of their Stokes vectors. Overall, this results in ${\sim}16$ gigabytes per wavelength and patch under the assumption of a flat detector at great distance. Considering how coarse the grid in the Stokes components is, it suggests that a proper treatment of polarization using the extended peel-off method is still questionable as it is currently limited by its enormous memory consumption. However, given the rapid development of computer hardware, it can be expected that a full treatment of polarization based on this approach will become feasible in the future.

\section{Summary and conclusions}
\label{sec:summary}

In this work, we studied the problems of high optical depths in MCRT simulations and various potential solutions. For this purpose, we adopted from \cite{2018ApJ...861...80C} the numerical setup of a three-dimensional infinite plane-parallel homogeneous slab and studied the transmission of radiation through it (see Sect. \ref{sec:slab_numerical_implementation}). To explore and extend possible solutions to this problem, we initially turned to a further simplified one-dimensional version of the slab experiment (Sect. \ref{sec:problems_high_optical_depth}). This analysis was based on a solution of a differential equation for photon distribution eigenstates in the slab and a decomposition of an arbitrary distribution in terms of these eigenstates. Doing so we derived an expression for the contribution of different scattering orders $n$ to the transmission spectrum through a slab in Sect. \ref{sec:app:application_to_a_mcrt_simulation}, the results of which are displayed in Fig. \ref{fig:oneD_slab_transmission}. We demonstrated that proper flux estimations for increasing optical depths involve not only the simulation of high scattering orders, but also wide ranges of different scattering orders. The scattering order problem is accompanied by a pathfinding problem;  the weight of the path of a photon package and its probability may strongly deviate from one another, in particular in the regime of high optical depths. We found that the albedo is a crucial variable that   determines the severity of the two problems. In particular, an increasing value of the albedo  increases the relevant range of scattering orders, and thus decreases the pace at which highly contributing photon packages need to travel through the model space. Based on these findings, and despite its performance boost, we argue that the \sas method will inevitably fail at high optical depths as it is designed to significantly reduce the scattering order of detected photon packages. 

In Sect. \ref{sec:methods} we identify the peel-off method as a promising candidate for solving the scattering order problem. When applied, the basic peel-off method repeatedly generates peel-off photon packages at every scattering point on the path of a representative photon package. Since peel-off photon packages are sent on an interaction-free path toward the detector, the detected set of peel-off photon packages comprises a range of scattering orders  $n\in\left\{ 0,1,\cdots,n_{\rm max}\right\}$, where $n_{\rm max}$ is the number of scattering events the original photon package has undergone before leaving the model space. In Fig. \ref{fig:SePS_vs_SPS_vs_CB} the \sps method, which combines the basic peel-off method with the \sas method, shows a slight improvement compared to the \sas method in the $\taumax=75$ slab experiment. Nonetheless, just like the \sas method, the \sps method also fails to quickly and properly calculate transmission spectra at high optical depths for a few reasons. First, the full range of relevant scattering orders has to be calculated, meaning that increasing values of $\taumax$ require increasing computational effort. Second, as seen in Fig. \ref{fig:oneD_slab_transmission}, detected photon packages of low scattering order become increasingly futile with higher optical depth. This means that a potentially high amount of computation time has to be spent on the calculation of mostly negligible scattering orders before stronger contributing scattering orders can even be calculated. 

For these reasons, we proposed a novel method that aims to solve the scattering order problem by taking into account environmental information of the path of a photon package:  the extended peel-off method. Rather than generating a peel-off photon package at a point of interaction and sending it to the detector, the interacting photon package is placed in the center of a homogeneous sphere, and a number of peel-off photon packages are then sent from the surface of the sphere to the detector. Subsequently, the original photon package is immediately moved to and emitted from the surface of that sphere. As a consequence, the extended peel-off method emits photon packages that are composed of contributions from multiple scattering orders. The emission properties of the sphere need to be precalculated and stored before this method can be used (see Sect. \ref{sec:extended_peel_off}). This approach tackles multiple problems at once. First, starting from its first application the extended peel-off method generates high scattering order photon packages, and thus the range of weakly contributing scattering orders is quickly surpassed and the range of highly contributing scattering orders reached. Second, the immediate displacement of the photon package skips the calculation of a potentially high number of scattering events during a MCRT simulation. Figure \ref{fig:SePS_vs_SPS_vs_CB} compares the \seps method, which combines the \sas method with the extended peel-off method, with the \sps method. Additionally, it shows results of the simple \seps method,   a variant in which the sphere is represented by a single peel-off photon package. The figure clearly shows that the \seps method  results in the most accurate transmission spectrum. Moreover, this simulation took only ${\sim}2$\,days, and   thus saved ${>}95\%$ of the computation time compared to the equivalent serial computation time reported by \cite{2018ApJ...861...80C}, while at the same time achieving a significantly lower error (see Sect. \ref{sec:results_and_discussion} for details).

Furthermore, we discuss the benefits of precalculating emission spectra compared to using spectra that are used for the MRW procedure. The two main reasons are first, that the directional emission properties of the surface of a sphere are important for a proper peel-off approach, and second, that MRW is not capable of simulating polarization and Mie-scattering. In Sect. \ref{sec:anisotropic_scattering_and_polarization}, the inclusion of anisotropic scattering or even polarization and Mie-scattering in the \seps method was described. While the inclusion of anisotropic scattering,   for instance HG scattering, is  possible, a full treatment of polarization is currently limited by the memory consumption of precalculated spectra and feasible only for rather coarsely sampled grids for the polarization states when using state-of-the-art computers. 

Furthermore, the limitations of the \seps method in terms of the transverse optical depth $\taumax$ have been studied. Figure \ref{fig:SePS_and_SPS_with_2dir_and_1dir} shows results of the \seps method for three different optical depths $\taumax\in\left\{ 75,150,300\right\}$. In this context we also tested an additional variant of the \seps method that deviates only in that it applies the Stretch method if and only if the photon package scatters toward the detector, hence the name forward \seps method. We find that the forward \seps method results in overall higher transmitted luminosities, as was expected, but that both variants have difficulties  properly simulating the spectra at $\taumax \geq 150$ when using $10^9$ photon packages. We note that it is possible to simulate higher values of optical depths at the cost of increased numbers of photon packages (i.e., with increased simulation time). However, in this regime of $\taumax$ the number of photon packages that is needed for a proper flux determination already reaches values that are too high to be handled with acceptable computational effort. We conclude that of all the tested methods, the \seps method showed the best performance for all studied transverse optical depth. This  makes it the preferred MCRT method that has been tested in the regime of high optical depths even though it is not fully capable of solving all problems that arise in this regime. The benefits it grants to MCRT simulations, however, significantly alleviate the issue of underestimated flux.  We also find that its success is mainly based on the coverage of scattering orders that it already provides  during the early stages of the life cycle of every photon package.

According to the findings in Sect. \ref{sec:problems_high_optical_depth}, which are supported by the results in Sect. \ref{sec:the_role_of_albedo}, it seems that the probability of generating significantly contributing paths decreases with $\taumax$ and with $A$, meaning that the portion of paths with negligibly low weight is found to increase. As a result, the chances of generating highly contributing photon package paths also decreases. This implies that any uncoordinated MC approach with a limited number of photon packages is out of reach of being able to sample the paths sufficiently well as the optical depth increases. Moreover, it suggests that the higher the optical depth, the more the choice of the path of a photon package has to be coordinated and the application of some form of biasing may be crucial for this undertaking. For the determination of flux in a setup of high optical depths, it thus seems inevitable to equip MCRT simulations with pathfinding techniques. We believe that the problem of high optical depths may pave the way for a new  bottom up  approach to MCRT simulations;  we believe that this new approach may lie in the use of artificial intelligence to guide photon packages on their way through the model space. Whether a method like reinforcement learning has its place in this undertaking remains to be seen. The simultaneous solution to the pathfinding and scattering order problem is the prerequirement to a full solution to the optical depth problem of MCRT simulations, and with this paper we have presented a method to solve the latter problem.





\begin{acknowledgements}
         We thank all the members of the Astrophysics Department Kiel for helpful discussions and remarks. 
         We acknowledge the support of the DFG priority program SPP 1992 "Exploring the Diversity of Extrasolar Planets (WO 857/17-1)".
\end{acknowledgements}

\bibliographystyle{aa} 
\bibliography{literature} 


\begin{appendix}

\section{Solution to the one-dimensional slab problem}
\label{sec:app:solution_one_dimensional_slab_problem}

In this section the problem of high optical depths of Monte Carlo radiative transfer (MCRT) simulations is studied. To this end, we consider the one-dimensional slab problem, derive its solutions in terms of eigenstates of the formulated problem, and    from the solution draw conclusions about the underlying difficulties a MCRT simulation encounters.

\subsection{One-dimensional slab problem}
\label{sec:app:setup_the_one_dimensional_slab_problem}
We consider a one-dimensional connected region of total optical depth $\tau_\text{tot}>0$  embedded in a vacuum. Let $v(x)$ be the distribution of photons, where $x\in \mathbb{R}$ is the (extinction) optical depth coordinate. Furthermore, we assume that $v(x)$ may only be non-zero for $x\in \left[ -b,b\right]$, where $b=\tau_\text{tot}/2$ defines the border of the region. In an iterative process, photons are released isotropically from their corresponding positions and experience their succeeding events of interaction at new positions. A portion of these photons, however, leave the region through one of its borders and do not interact any longer. The randomly traversed individual distance $\Delta x$ that a photon travels follows a distribution $p(\Delta x)$ given by 
\begin{equation}
p(\Delta x) = \frac12 \exp\left(-\abs{\Delta x} \right).
\end{equation}
At the location of interaction a photon package either scatters and continues its path through the medium or is absorbed. The probability for scattering and absorption is given by $A$ and $1-A$, respectively, where $A$ is the albedo. The albedo is defined as the fraction between the scattering cross section $C_\text{sca}$ and the extinction cross section $C_\text{ext}$. The expected resulting distribution of interacting photons $\tilde{v}(x)$ can then be written as follows:
\begin{equation}
\tilde{v}(x) = \frac12 \int_{-b}^{b} dy\,v(y) \exp\left( - \abs{x-y} \right).
\label{eq:general_equation}
\end{equation}
In the following we derive eigenstates of this operation and formula to expand $v(x)$ in terms of these eigenstates. 

\subsection{Derivation of eigenstates}
\label{sec:app:derivation_of_eigenstates}
Eigenstates $v(x)$ of the operation in Eq. \eqref{eq:general_equation} with their corresponding eigenvalues $\lambda$ satisfy the equation
\begin{equation}
\lambda v(x) = \frac12 \int_{-b}^{b} dy\,v(y) \exp\left( - \abs{x-y} \right).
\label{eq:general_eigenvalue_eq}
\end{equation}
By writing the exponential function in Eq.  \eqref{eq:general_eigenvalue_eq} as 
\begin{equation}
\exp\left( - \abs{x-y} \right) = \text{H}\left( x-y\right) \exp\left( y-x \right) + \text{H}\left( y-x\right)\exp\left( x-y \right),
\end{equation}
where $\text{H}(x)$ is the Heaviside step function, and taking the second derivative of Eq. \eqref{eq:general_eigenvalue_eq}, we obtain the following second-order differential equation:
\begin{equation}
\lambda v'' + \left( 1 - \lambda \right) v = 0.
\label{eq:differential_eq}
\end{equation}
Here $v''$ denotes the second derivative of $v$ with respect to $x$.  For $0<\lambda<1$, the solutions for this common differential equation are given by the anti-symmetric function $\sin\left(\sigma x\right)$ and the symmetric function $\cos\left(\sigma x\right)$, where 
\begin{equation}
\sigma^2 = \frac{1}{\lambda} - 1.
\label{eq:lambda_sigma}
\end{equation}
However, when inserting these solutions into Eq. \eqref{eq:general_eigenvalue_eq} we find conditions for $\sigma$ that have to be satisfied for these eigenstates. The condition for the symmetric and the anti-symmetric eigenstates are 
\begin{align}
&\cos\left( \sigma b\right) = \sigma\, \sin\left( \sigma b\right)& (\text{symmetric states}),\\
&\sin\left( \sigma^* b\right) =- \sigma^*\, \cos\left( \sigma^* b\right)& (\text{anti-symmetric states}),
\end{align}
where the asterisk  indicates that is belongs to the anti-symmetric solution. We define the sets of positive solutions for $\sigma$ to these conditions as $\Sigma$ and $\Sigma^*$, respectively the symmetric and anti-symmetric solutions, which are thus given by
\begin{align}
\begin{split}
\Sigma &= \left\{ \sigma_i  \,\middle\vert\, i \in \mathbb{N}_{>0} \,\wedge\, i-1<\frac{\sigma_i b}{\pi} <i  \,\wedge\, \cdots \right.\\ &
\left. \mathrel{\phantom{\frac12 \quad}} \cdots \wedge\, \cos\bigl( \sigma_i b\bigr) = \sigma_i\, \sin\bigl( \sigma_i b\bigr)  \right\}
\end{split}
\end{align}
and
\begin{align}
\begin{split}
\Sigma^* &= \left\{ \sigma_i^*  \,\middle\vert\, i \in \mathbb{N}_{>0} \,\wedge\,  i-\frac12<\frac{\sigma_i^* b}{\pi} <i+\frac12 \,\wedge\, \cdots \right.\\ &
\left. \mathrel{\phantom{\frac12 \quad}} \cdots \wedge\, \sin\bigl( \sigma_i^*  b\bigr) =- \sigma_i^* \, \cos\bigl( \sigma_i^*  b\bigr)  \right\}. 
\end{split}
\end{align}
The values of $\sigma_i \in \Sigma$ and $\sigma_i^* \in \Sigma^*$  increase for increasing $i \in \mathbb{N}_{>0}$, and the sets $\Sigma$ and $\Sigma^*$ are disjoint. Moreover, we define 
\begin{equation}
\lambda_i = \frac{1}{1+{\sigma_i}^2} \,\wedge \, \lambda_i^* = \frac{1}{1+{\sigma_i^*}^2} \,\, \forall i \in \mathbb{N}_{>0}.
\label{eq:lambda_i}
\end{equation}
Let $\mathcal{F}$ be the set of eigenstates
\begin{equation}
\mathcal{F} = \left\{ \cos\bigl( \sigma_i x\bigr) \,\middle\vert\, i \in \mathbb{N}_{>0}  \right\} \cup \left\{ \sin\left( \sigma_i^* x\right) \,\middle\vert\, i \in \mathbb{N}_{>0}  \right\}.
\end{equation}
We define the inner product $\left\langle \cdot\,,\cdot \right\rangle$ as
\begin{equation}
\left\langle f\,,g \right\rangle = \int_{-b}^b dx\,\overline{f(x)} g(x),
\end{equation}
with $\overline{f(x)} \doteq f(x) / (b+\lambda_f)$ for all $f\in\mathcal{F}$, where $\lambda_f$ is the eigenvalue corresponding to $f$ obtained by Eq. \eqref{eq:lambda_i}. Using these definitions we find that the set $\mathcal{F}$ forms an orthonormal system of eigenstates: 
\begin{equation}
  \forall f,g \in \mathcal{F}:\, \left\langle f\,,g \right\rangle = 
  \begin{cases} 
      1 &\text{if }f=g\\
      0 &\text{else}
   \end{cases}.
\end{equation}
Next, we expand $v(x)$ in terms of the eigenstates $\mathcal{F}$ as 
\begin{equation}
v(x) = \sum_{k=1}^\infty a_k \cos\Bigl( \sigma_k x\Bigr) + b_k \sin\Bigl( \sigma_k^* x\Bigr),
\label{eq:expansion_in_eigenstates}
\end{equation} 
where the coefficients are given by 
\begin{equation}
a_k = \Bigl\langle \cos\Bigl( \sigma_k x\Bigr) \mathrel{\vphantom{\sigma_k^*}},v(x) \Bigr\rangle \quad \wedge \quad b_k = \left\langle \sin\left( \sigma_k^* x\right) \,,v(x) \right\rangle.
\label{eq:app:a_k_b_k}
\end{equation}

\subsection{Evolution of eigenstates}

Let $v_0(x)$ be the initial non-zero distribution of photons. After launch, the resulting distribution of interacting photons $\tilde{v}_0(x)$ is partly absorbed, while the remaining part $v_1(x) \doteq A \tilde{v}_0(x)$ continues its path through the medium. The distribution of launching photons after the $n$-th event of scattering satisfies
\begin{equation}
v_n(x) = A \tilde{v}_{n-1}.
\end{equation}
Let $a_k$ and $b_k$ with $k \in \mathbb{N}_{>0}$ be the expansion coefficients of $v_0$.  Subsequently, the distribution $v_n$ can be written as
\begin{equation}
v_n(x)= A^n \sum_{k=1}^\infty {\lambda_k}^n a_k \cos\Bigl( \sigma_k x\Bigr) + {\lambda_k^*}^n b_k \sin\Bigl( \sigma_k^* x\Bigr),
\end{equation}
where $\lambda_k$ and $\lambda_k^*$ are the eigenvalues corresponding to $\sigma_k$ and $\sigma_k^*$, respectively, which are linked by Eq. \eqref{eq:lambda_sigma}. Given that all corresponding eigenvalues satisfy $0<\lambda<1$, we find that the absolute value of the amplitude of every mode of the expansion decreases exponentially with increasing number n and vanishes only in the limit $n\rightarrow\infty$. Additionally, all eigenvalues are pairwise different; in particular, they satisfy the relation $\lambda_k > {\lambda_k}^* > \lambda_{k+1} > {\lambda_{k+1}}^* $ for all $k \in \mathbb{N}_{>0}$. Thus, any non-zero distribution $v(x)$ has a dominant (i.e., highest) eigenvalue $\lambda_{\rm max}$, that either corresponds to a dominant $\sigma_{\rm min} \in \Sigma$ or $\sigma_{\rm min} \in \Sigma^*$.  Therefore, as $n \rightarrow \infty$, the distribution asymptotically converges: 
\begin{equation}
  v_n(x) \xrightarrow[n \to\infty]{} A^n {\lambda_{{\rm max}}^n} \cdot
  \begin{cases} 
      \cos\left( \sigma_{\rm min} x\right) &\text{if }\sigma_{\rm min} \in \Sigma\\
      \sin\left( \sigma_{\rm min} x\right) &\text{if }\sigma_{\rm min} \in \Sigma^*.
   \end{cases}
\end{equation}
Furthermore, in the case of a non-zero non-negative initial distribution, the dominant eigenvalue is $\lambda_1$. Therefore, the shape of this photon distribution converges to the shape of the dominant eigenstate $\cos\left( \sigma_1 x\right)$, independent of the  specifics of the initital distribution, which is positive over the full domain of the slab\footnote{Interestingly, this result can also be obtained by discretizing the problem and describing the evolution of an arbitrary $m$-dimensional initial photon distribution by a $m\times m$ transition matrix $T$. Due to the simplicity of the problem, the matrix is a Toeplitz matrix that is positive definite, which is a consequence of its dominant diagonal entries. As a result, due to the Perron-Frobenius theorem, $T$ has a positive largest (dominant) eigenvalue with a corresponding one-dimensional eigenspace, which is spanned by a positive eigenvector.}.

\subsection{Observer}
\label{sec:app:observer}
In this subsection we derive the relevant equations that let us  study the problem of high optical depth and help us to find tools to overcome this problem. To do this we place an observer at $x\rightarrow \infty$ who detects photons that leave the optically thick region through its border at $x=b$. Given a distribution of launching photons $v_0(x)$, the measurement of observed photons during a single step is given by
\begin{equation}
I(v_0) = \frac12 \int_{-b}^{b} dy\,v_0(y) \exp\left( y-b \right).
\label{eq:general_observed_intensity}
\end{equation}
Comparing Eq. \eqref{eq:general_observed_intensity} with \eqref{eq:general_equation} we find
\begin{equation}
I(v_0) = \tilde{v}_1(b),
\end{equation}
meaning that after a single step of photon transfer the measurement of the observer equals the photon density at the border of the region that is facing the observer, prior to its interaction with the medium. This equation holds for arbitrary distributions $v(x)$. Thus, after the $n$-th event of interaction and assuming $A\neq 0$, the measurement $I^{(n)}$ can be written as
\begin{equation}
I^{(n)} = \frac{v_{n+1}(b)}{A}. 
\end{equation}
Using the expansion coefficients $a_k$ and $b_k$ of $v_0(x)$, we find the general expression of the measurement to be given by
\begin{align}
I^{(n)} &= A^n \sum_{k=1}^\infty {\lambda_k}^{n+1} a_k \cos\Bigl( \sigma_k b\Bigr) + {\lambda_k^*}^{n+1} b_k \sin\Bigl( \sigma_k^* b\Bigr) \\
           &= A^n \sum_{k=1}^\infty \left( -1 \right)^{k+1}\left( {\lambda_k}^{n+1} \sqrt{1-\lambda_k} \,a_k+ {\lambda_k^*}^{n+1} \sqrt{1-\lambda_k^*}\, b_k\right).
\end{align}
For  equation (A.24) we made use of Eqs. \eqref{eq:app:cos_sigma} and \eqref{eq:app:sin_sigma_star}. The measurement of the observer increases with the number of considered interactions to a partial sum $I^{(\leq n)}$ given by
\begin{align}
I^{(\leq n)} &= \sum_{m=0}^{n} I^{(m)}  \\
\begin{split}
= \sum_{k=1}^\infty \left( -1 \right)^{k+1}\Biggl( \lambda_k\, \frac{1-\left( A\lambda_k\right)^{n+1}}{1-A\lambda_k} \sqrt{1-\lambda_k} \,a_k+ \cdots \\
\quad \cdots + \lambda_k^*\, \frac{1-\left( A\lambda_k^*\right)^{n+1}}{1-A\lambda_k^*} \sqrt{1-\lambda_k^*}\, b_k\Biggr).
\end{split}
\label{eq:app:I_n}
\end{align}
Since $\abs{A\lambda_k}<1$ and  $\abs{A\lambda_k^*}<1$, we can take the limit of the partial sums and arrive at the total measured quantity $I^\text{tot}$: 
\begin{equation}
I^\text{tot} = \sum_{k=1}^\infty \left( -1 \right)^{k+1}\Biggl(\frac{ \lambda_k \sqrt{1-\lambda_k} }{1-A\lambda_k} \,a_k +  \frac{\lambda_k^*\sqrt{1-\lambda_k^*}}{1-A\lambda_k^*} \, b_k\Biggr).
\label{eq:observed_total}
\end{equation}

\subsection{Application to a MCRT simulation}
\label{sec:app:application_to_a_mcrt_simulation}
In this section we apply the results from Sect. \ref{sec:app:observer} to the case of a MCRT simulation, in particular, we derive a solution for the transmitted energy for the case of an one-dimensional slab. 

During a MCRT simulation isolated photon packages originate from different points in the model space and travel through it. The photon distribution thus corresponds to a delta distribution that  launches at a position $x=a$:
\begin{equation}
v_0(x) = E\,\delta (x-a).
\label{eq:single_photon}
\end{equation}
Here $E$ is the energy carried by the photon package. It then generally interacts multiple times before leaving its cell. As a consequence, on average, the energy of the photon package spreads over the entire region according to 
\begin{equation}
v^\text{tot} = \sum_{n=0}^\infty v_n(x).
\end{equation}
Depending on the initial position $a \in \left[-b, b \right]$, the energy that the observer detects can be calculated using Eq. \eqref{eq:observed_total}. Assuming an albedo of $A=1$ and using the relations from Sect. \ref{sec:app:useful_relations}, the total observed intensity is given by
\begin{equation}
I^\text{tot} = \frac{E}{2} \left( 1 + \frac{a}{b+1}\right),
\label{eq:app:I_tot}
\end{equation}
which in the case of $a=0$ equals exactly $E/2$; in the case $a=b$ it reaches its maximum value, which asymptotically converges to $E$ as $b\rightarrow \infty$; and finally in the case of $a=-b$ it reaches its minimum value of $E/(2b+2)$, which vanishes as $b\rightarrow \infty$. 

\paragraph*{Transmission through a one-dimensional slab:} 
\label{sec:app:transmission_trhough_one_dim_slab}
A slab, as described in Sect. \ref{sec:app:setup_the_one_dimensional_slab_problem}, that is illuminated from the left (i.e., from $x\rightarrow -\infty$) can be described by the initial distribution $v_0(x) = 2\delta(x+b)$. As a result, the energy that enters the slab is normalized to 1. 
Using Eqs. \eqref{eq:app:a_k_b_k} and \eqref{eq:app:I_n} we arrive at the contribution $I^{(n)}$ to the total transmitted given by\begin{equation}
I^{(n)}\doteq A^n\, I^{(n)}_{A=1}=2 A^n \sum_{k=1}^\infty \frac{\lambda_k^{n+1}\Bigl( 1-\lambda_k \Bigr)}{b+\lambda_k} - \frac{{\lambda_k^*}^{n+1}\Bigl( 1-{\lambda_k^*} \Bigr)}{b+{\lambda_k^*}}.
\label{eq:app:transmission_slab}
\end{equation} 
We note that $I^{(n)}$ can also be calculated by the $n$-dimensional integral
\begin{equation}
I^{(n)} = \left(\frac{A}{2}\right)^n \int_{\left[ 0,\tau_{\rm tot}\right]^n} d\tau_1\cdots d\tau_n \, \exp\left\{ -\sum_{k=0}^n \abs{\tau_{k+1}-\tau_k} \right\},
\label{eq:app:I_n_integralform}
\end{equation}
with $\tau_0\doteq0$ and $\tau_{n+1}\doteq \tau_{\rm tot}$. The sum in Eq. \ref{eq:app:transmission_slab} tends to converge slowly for low scattering orders $n$, which is a consequence of the slow decrease in the eigenvalues $\lambda_k$ with $k$. The calculation of $I^{(n)}$ based on its integral form, however, is simple in this case, while becoming impractical for increasing scattering order. Therefore, the integral form can be solved analytically up to a certain $n$ and used a measure for convergence of the sum in Eq. \ref{eq:app:transmission_slab}.

The overall observed energy of photons is the sum of all contribution $I^{(n)}$. The average number of scattering events a photon package has undergone before its detection ($N^\text{mean}$) depends on the optical depth and is given by
\begin{equation}
N^\text{mean} = \frac1{I^\text{tot}}\sum_{n=0}^{\infty} n\,I^{(n)} 
\label{eq:app:n_mean}
.\end{equation}
It is obvious that $N^\text{mean}$ increases with the albedo, and thus  in the case of $A=1$  reaches its highest value ($N^\text{mean}_{A=1}$). Using equations from Sect. \ref{sec:app:useful_relations}, it is given by
\begin{equation}
N^\text{mean}_{A=1} = \frac{2b^3+6b^2+3b}{3b+3}.
\label{eq:app:N_mean}
\end{equation}
For $0<b\ll 1$ we find that $N^\text{mean}_{A=1} \sim b$ and for high optical depths (i.e., $b\gg 1$,  $N^\text{mean}_{A=1} \sim 2b^2/3$). Thus, in order to simulate high optical depths properly, MCRT simulations are required to simulate photon packages of increasingly high scattering order.

\subsection{Useful relations}
\label{sec:app:useful_relations}
When using the results from Sect. \ref{sec:app:derivation_of_eigenstates} and expanding $v(x)\in\left\{ 1,x,x^2,x^3\right\}$
in terms of the eigenstates in $\mathcal{F}$, we can derive multiple relations, some of which are   listed in the following.
We note that specific variable names (i.e., $\sigma$,$\lambda$) marked with an asterisk   correspond to anti-symmetric eigenstates, while those without the asterisk   correspond to symmetric eigenstates. 
\begin{alignat}{2}
&\cos\left( \sigma_n b\right) && = (-1)^{n+1} \sqrt{1-\lambda_n}  \label{eq:app:cos_sigma}\\
&\sin\left( \sigma_n b\right) && = (-1)^{n+1} \sqrt{\lambda_n} \\
&\cos\left( {\sigma_n}^* b\right) && = (-1)^{n} \sqrt{{\lambda_n}^*} \\
&\sin\left( {\sigma_n}^* b\right) && = (-1)^{n+1} \sqrt{1-{\lambda_n}^*} \label{eq:app:sin_sigma_star}\\
&\sum_{n=1}^\infty \frac{\sin\left( \sigma_n b\right) }{\sigma_n}\frac{\cos\left( \sigma_n a\right)}{b+\lambda_n} && = \frac{1}{2} \quad \forall a \in \left[ -b,b  \right]\\
&\sum_{n=1}^\infty \frac{\lambda_n}{b+\lambda_n} && = \frac{1}{2} \\
&\sum_{n=1}^\infty \frac{\left( -1 \right)^{n+1}}{b+\lambda_n} \frac{\lambda_n}{\sqrt{1-\lambda_n}}  && = \frac{1}{2} \\
&\sum_{n=1}^\infty \frac{1}{b+\lambda_n}\frac{\lambda_n^2}{1-\lambda_n} && = \frac{b}{2} \\
&\sum_{n=1}^\infty \frac{\cos\left( \sigma_n^* b\right) }{\sigma_n^*}\frac{\sin\left( \sigma_n^* a\right)}{b+\lambda_n^*} && = -\frac{1}{2}\frac{a}{b+1} \quad \forall a \in \left[ -b,b  \right]\\
&\sum_{n=1}^\infty \frac{\lambda_n^*}{b+\lambda_n^*} && =\frac{1}{2}\frac{b}{b+1} \\
&\sum_{n=1}^\infty \frac{\sin\left( \sigma_n b\right) }{\sigma_n^3}\frac{\cos\left( \sigma_n a\right)}{b+\lambda_n} && = \frac{1}{4}\left( b^2+2b - a^2 \right) \quad \forall a \in \left[ -b,b  \right]\\
&\sum_{n=1}^\infty \frac{\left( -1\right)^{n+1}}{b+\lambda_n} \frac{\lambda_n^2}{\left(1-\lambda_n\right)^{\frac{3}{2}}} && = \frac{b}{4}(b+2)\\
&\sum_{n=1}^\infty \frac{\cos\left( {\sigma_n}^* b\right) }{{\sigma_n^*}^3}\frac{\sin\left( \sigma_n^* a\right)}{b+\lambda_n^*} && = \frac{a^3\left( 1+b\right) - ab^2\left( b+3 \right)}{12 \left( 1+b\right)^2} \quad \forall a \in \left[ -b,b  \right]\\
&\sum_{n=1}^\infty \frac{1}{b+\lambda_n^*}\frac{{\lambda_n^*}^2}{1-\lambda_n^*} && = \frac{1}{6} \frac{b^3}{\left( 1+b \right)^2}
\end{alignat}

\section{Method descriptions and transmission spectra}

\begin{table*}[!h]
\centering
\begin{tabular}{ m{2.5cm} | m{12cm} | m{2cm}  } 
\textbf{Method}& \textbf{Description}& \textbf{Reference} \\ 
\hline
\hline
Split & Whenever a photon package interacts it is split into two photon packages, one of which is absorbed and the other one scatters. The latter one carries a weight that is reduced according to the scattering probability.& Sect. \ref{sec:slab_numerical_implementation}\\  \hline

Stretch & The path length determination of a photon package is modified in favor of longer interaction-free path segments. The traversed path length is chosen according to the PDF for composite-biasing. In order to compensate for this modification the weight of the photon package needs to be readjusted. & Sect. \ref{sec:slab_numerical_implementation} \\  \hline

Basic peel-off & At every point of interaction an additional peel-off photon package is generated that travels interaction-free toward the detector. Its weight is chosen according to the scattering probability in that particular direction. The hereafter detected luminosity is additionally reduced due to the optical depth encountered on the way to the detector. & Sect. \ref{sec:basic_peel_off}\\  \hline

Extended peel-off & A homogeneous sphere is defined with the point of interaction of the photon package as its center. The interacting photon package is immediately emitted from the surface of that sphere and its weight is reduced according to the corresponding emission spectrum of that sphere. Additionally, multiple peel-off photon packages are launched interaction-free from different locations on the surface of the sphere to the detector. &Sect. \ref{sec:extended_peel_off}\\  \hline
\hline

\sas& A combination of the Split method and and the Stretch method is applied. & Sect. \ref{sec:slab_numerical_implementation}\\  \hline
\sps& A combination of the Split method, the basic peel-off method, and the Stretch method is applied. &Sect. \ref{sec:extended_peel_off}\\  \hline
\seps&  A combination of the Split method, the basic peel-off method, the extended peel-off method, and the Stretch method is applied. In particular, the Split method and the Stretch method are used during every interaction of a photon package, while the extended peel-off method is only used during interactions that allow for its usage and the basic peel-off method in the case that the extended peel-off method cannot be used. &Sect. \ref{sec:extended_peel_off}\\  \hline
\hline

Forward \sps& A version of the \sps method is applied, which differs from the \sps method only in its use of the Stretch method. In particular, during its application the Stretch method is used if and only if the direction of the photon package after scattering points toward the detector, i.e., in the forward direction. & Sect. \ref{sec:role_of_peel_off_on_mcrt}\\  \hline
Forward \seps& A version of the \seps method is applied, which differs from the \seps method only in its use of the Stretch method. In particular, during its application the Stretch method is used if and only if the direction of the photon package after scattering points toward the detector, i.e., in the forward direction. & Sect. \ref{sec:role_of_peel_off_on_mcrt}\\  \hline
Simple \seps& A version of the \seps method is applied, which differs from the \seps method only in its use of the extended peel-off method. Instead of sending multiple peel-off photon packages from different locations on the surface of the sphere to the detector, only a single peel-off photon package is generated and sent to the detector. & Sect. \ref{sec:extended_peel_off}\\  
\end{tabular}
\caption{Overview of the MCRT methods  used in this paper. Details are provided in the referenced sections.}
\label{tab:overview_mcrt_methods}
\end{table*}


   \begin{figure*}[!h]
   \centering
   \includegraphics[width=0.45\textwidth]{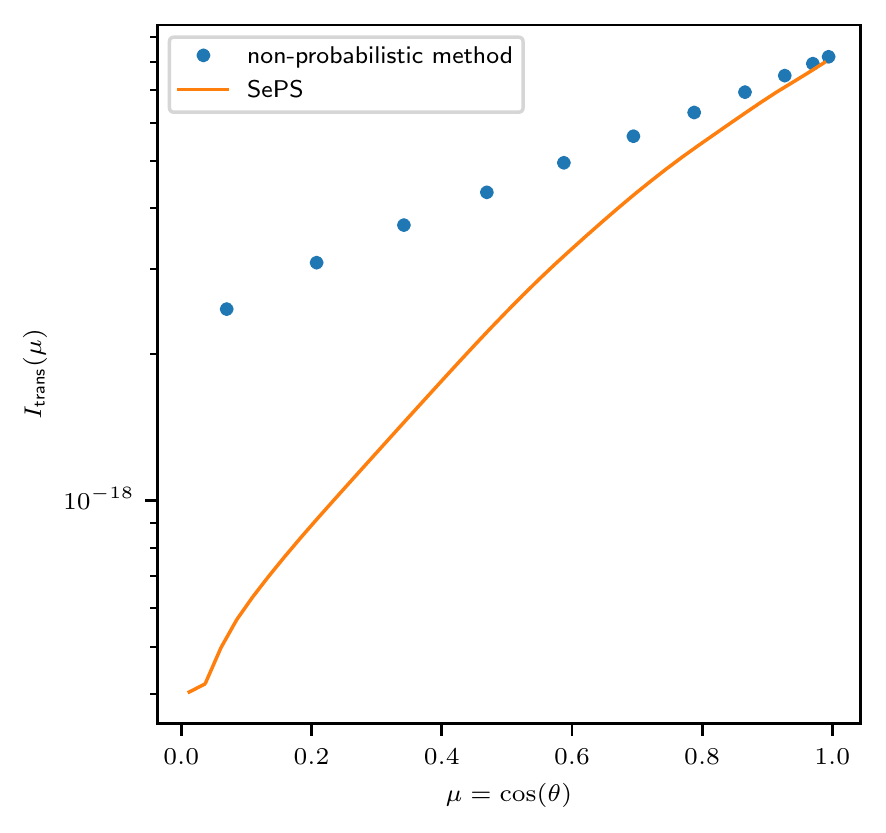}
   \caption{Results for the transmitted intensity $I_{\rm trans}$ through an infinite plane-parallel slab with an albedo of $A=0.9$ and a transverse optical depth of $\taumax=75$, where $\mu = \cos(\theta)$ is the cosine of the penetration angle $\theta$. The \seps method (solid line) is compared against a result from a non-probabilistic numerical method (dotted line). For details, see Sect. \ref{sec:the_role_of_albedo}.}
              \label{fig:SePS_for_largerA}
   \end{figure*}

   \begin{figure*}[!h]
   \centering
   \includegraphics[width=0.75\textwidth]{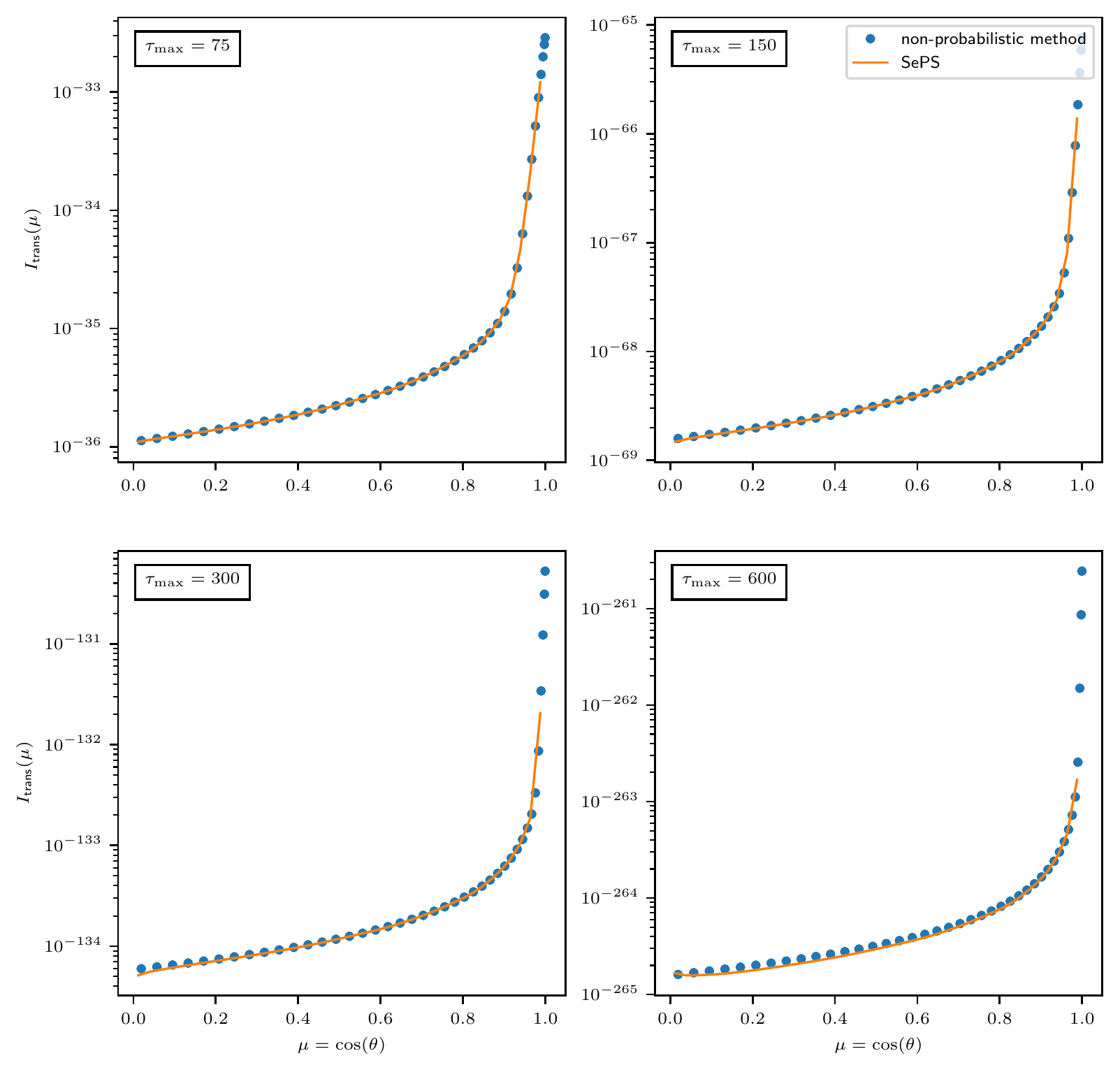}
   \caption{Results for the transmitted intensity $I_{\rm trans}$ through an infinite plane-parallel slab with an albedo of $A=0.9$ and for transverse optical depths $\taumax\in\left\{150,300,600\right\}$, where $\mu = \cos(\theta)$ is the cosine of the penetration angle $\theta$. The \seps method (solid lines) is compared against results from a non-probabilistic numerical method (dotted lines). For details, see Sect. \ref{sec:the_role_of_albedo}.}
              \label{fig:SePS_for_smallerA}
   \end{figure*}

\end{appendix}

\end{document}